%% file: main.tex
\documentclass[runningheads]{llncs}

\usepackage{eccv}

\usepackage{eccvabbrv}

\usepackage{graphicx}
\usepackage{booktabs}
\usepackage{tabularx}
\usepackage{multirow}
\usepackage{siunitx}
\usepackage{float}
\usepackage{comment}

\usepackage[accsupp]{axessibility} 

\usepackage{hyperref}

\usepackage{orcidlink}

\begin{document}

\input{macros.tex}


\title{\ours: \underline{D}iffusion Rendering for \underline{R}elightable \underline{Ex}pressive Avatars} 

\author{
Timo Teufel\inst{1}\and
Xilong Zhou\inst{1}\and
Umar Iqbal\inst{2}\and
Jan Kautz\inst{2}\and
Marc Habermann\inst{1}\and
Vladislav Golyanik\inst{1}\and
Christian Theobalt\inst{1}\vspace{-5pt}}

\authorrunning{Teufel et al.}

\institute{$^{1}$Max Planck Institute for Informatics, SIC \quad $^{2}$NVIDIA}

\maketitle

\input{Main_Sections/00_abstract}
\input{Main_Sections/10_introduction}
\input{Main_Sections/20_related_work}
\input{Main_Sections/30_method}
\input{Main_Sections/40_results}

\input{Main_Sections/45_Limitations}
\input{Main_Sections/50_conclusion}

\clearpage

%
%
\bibliographystyle{splncs04}
\bibliography{main}


\clearpage

\title{\ours: \underline{D}iffusion Rendering for \underline{R}elightable \underline{Ex}pressive Avatars}
\subtitle{Supplementary Material}

\author{
Timo Teufel\inst{1}\and
Xilong Zhou\inst{1}\and
Umar Iqbal\inst{2}\and
Jan Kautz\inst{2}\and
Marc Habermann\inst{1}\and
Vladislav Golyanik\inst{1}\and
Christian Theobalt\inst{1}\vspace{-5pt}}

\authorrunning{Teufel et al.}

\institute{$^{1}$Max Planck Institute for Informatics, SIC \quad $^{2}$NVIDIA}

\maketitle

\renewcommand{\thesection}{\Alph{section}}
\setcounter{section}{0}
\setcounter{figure}{0}
\setcounter{table}{0}
\renewcommand{\thefigure}{A\arabic{figure}}
\renewcommand{\thetable}{A\arabic{table}}

\input{Supp_Sections/00_overview}
\input{Supp_Sections/01_data_capture}
\input{Supp_Sections/02_additional_results}
\input{Supp_Sections/03_generalization}
\input{Supp_Sections/05_shift}
\input{Supp_Sections/06_ethical_discussion}

\end{document}

%% file: macros.tex
\newcommand{\xz}[1]{\textcolor{red}{{XZ:#1}}}
\newcommand{\vg}[1]{\textcolor{blue}{{VG: #1}}}
\newcommand{\mhc}[1]{\textcolor{magenta}{{MH: #1}}}
\newcommand{\ui}[1]{\textcolor{green}{{UI: #1}}}
\newcommand{\tet}[1]{\textcolor{orange}{{TT: #1}}}

\newcommand{\ours}{\textit{D-Rex}}
\newcommand{\ourss}{\textit{D-Rexs}}

%% file: Main_Sections/00_abstract.tex
\vspace{-20pt}
\begin{figure}
    \centering
    \includegraphics[width=\textwidth]{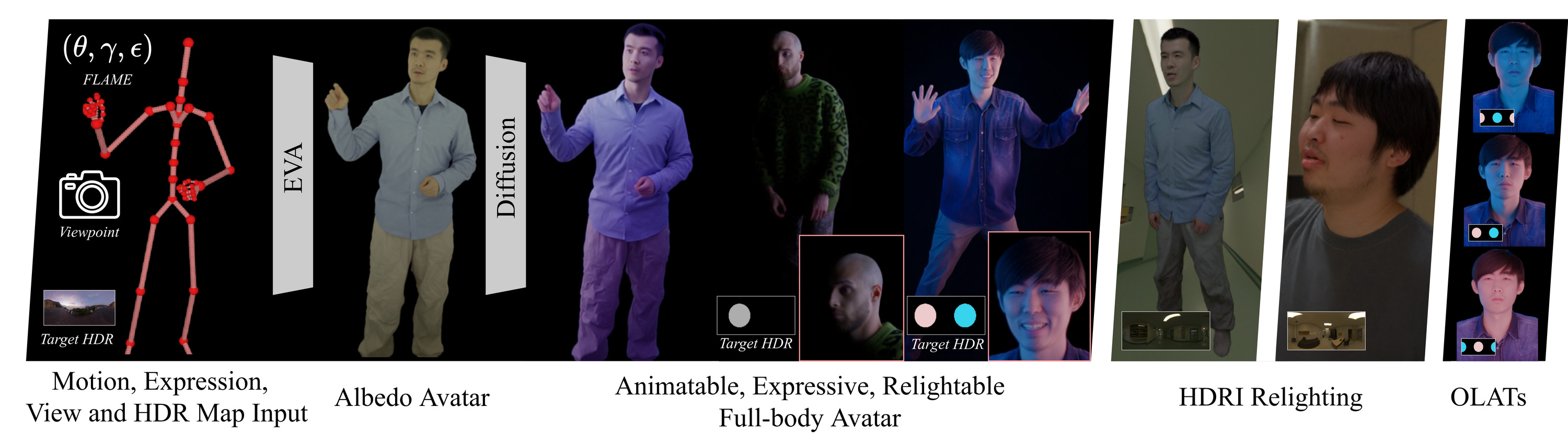}
    \vspace{-15pt}
    \caption{Given 3D body motion, facial expressions, a target viewpoint, and a High Dynamic Range (HDR)
  environment map, \ours{} renders photorealistic, relightable images of the person under novel views
  and illuminations. We show that fine-tuning a pre-trained video diffusion relighting model solely on single-image flat-lit to relit frame pairs is sufficient to achieve convincing view and temporal consistency across frames.}
    \label{fig:teaser}
\end{figure}
\vspace{-5pt}
\vspace{-30pt}
\begin{abstract}

We present \ours, a person-specific framework for photorealistic, relightable, expressive, and animatable full-body human avatars with free-viewpoint rendering.
Existing methods for relightable full-body avatars rely on explicit 3D intrinsic decomposition with analytic reflectance models, which require accurate geometry registration and careful optimization to capture realistic light transport effects.
This tight coupling of relighting with avatar modeling has hindered expressiveness: to our knowledge, no existing method demonstrates strong facial animation alongside relighting, limiting applicability in telepresence, gaming, and virtual production.
We propose to decouple relighting entirely from avatar modeling by treating it as an image-space post-process: a learned translation from flat-lit, albedo-like renderings to a target HDR illumination.
To this end, we leverage the strong generative prior of a pre-trained video diffusion relighting model, fine-tuned via LoRA on paired flat-lit and relit frames captured in a light stage.
The flat-lit driving frames are produced by an independent expressive full-body avatar framework trained under white-light conditions, requiring no modification to support relighting, making \ours~directly applicable to any white-light avatar system.
We demonstrate that \ours~enables view- and temporally consistent relighting while faithfully preserving expressive motion and fine-grained facial detail, outperforming physically-based relightable avatar baselines.
Project page: \href{https://vcai.mpi-inf.mpg.de/projects/DRex/}{https://vcai.mpi-inf.mpg.de/projects/DRex/}
\keywords{Expressive Human Avatar \and Full-body Relighting \and Video Diffusion}
\end{abstract}

%% file: Main_Sections/10_introduction.tex
\section{Introduction}

\textit{Animatable, relightable full-body human avatars} aim to render a specific person
photorealistically under arbitrary lighting, driven by pose and facial expression, from any viewpoint.
Such avatars have broad applications in telepresence, character animation for games, virtual production,
and immersive VR/AR.

Despite remarkable progress in non-relightable full-body avatars, achieving high-resolution
photorealism~\cite{shao2025degas, pang2024ash, junkawitsch2025eva}, expressive facial
animation~\cite{shao2025degas, junkawitsch2025eva}, and dynamic clothing~\cite{habermann2021ddc,
pang2024ash, junkawitsch2025eva}, jointly supporting relighting, animation, and expressiveness remains
an open challenge. Existing relightable methods rely on explicit 3D intrinsic decomposition with analytic
reflectance models. Recovering accurate, pose-dependent intrinsics from multi-view video is highly
ill-posed, causing most methods to produce limited light transport quality even without expression
control~\cite{Wang2024IntrinsicAvatar, chen2024meshavatarlearninghighqualitytriangular,
iqbal2022ranarelightablearticulatedneural, Jiang2025DNF}. The recent work of Wang et
al.~\cite{WangARXIV2025fullbodyrgca} achieves visually compelling results through physically-based
rendering, but requires highly accurate geometry for training, and does not demonstrate a wide range of
facial expressions.

Meanwhile, a parallel line of work bypasses analytic reflectance modeling via \textit{image-space
relighting}: directly learning to translate images from one lighting condition to another. Methods for
portrait relighting~\cite{He2024DifFRelight, wang2025comprehensive, pandey2021total, Sun2019SingleImage,
zhang2021neuralvideoportraitrelighting, kim2024switchlight, mei2025lux} and general scene
relighting~\cite{fang2025vrgbxvideoeditingaccurate, liang2025DiffusionRenderer, he2025unirelight} achieve
impressive perceptual quality, but are not designed for controllable full-body avatars. Most relevant to
our work, He et al.~\cite{He2024DifFRelight} demonstrate diffusion-based person-specific relighting for
facial performances, but require a custom network and expensive One-Light-At-a-Time
(OLAT)~\cite{debevec2000acquiring} compositing for general HDR relighting, and their method does not
scale to full-body humans.

In this work, we present \ours, a framework that bridges both lines of work: combining high-quality
expressive full-body avatar rendering with powerful image-space diffusion relighting, bypassing
intrinsic decomposition entirely. Our key insight is that relighting can be cleanly decoupled from avatar
modeling by treating it as an image-space post-process, a learned translation from flat-lit,
albedo-like renderings to a target HDR illumination. We capture interleaved flat-lit and HDR-illuminated
multi-view frames in a light stage, covering a wide range of poses and expressions. We then train an
expressive full-body avatar~\cite{junkawitsch2025eva} (EVA) on the flat-lit frames to produce
albedo-like renderings for arbitrary pose and expression, and fine-tune a pre-trained video diffusion
relighting model~\cite{liang2025DiffusionRenderer} (Cosmos DiffusionRenderer) via LoRA on the paired
flat-lit/relit frames. Because the relighting step is fully decoupled, the expressive avatar requires no
modification to support relighting, making our approach directly applicable to any white-light avatar
system. With only lightweight fine-tuning, \ours~produces perceptually realistic, view- and temporally
consistent renderings that outperform our physically-based relightable avatar baseline. A demonstration
of the visual quality achieved by our method is shown in~\cref{fig:teaser}.

The main technical contributions of this work are:
\begin{itemize}
    \item[$\circ$] \ours, the first diffusion-based relightable full-body avatar framework, enabling
    photorealistic, view-consistent rendering under novel poses, facial expressions, viewpoints, and
    illuminations [\cref{sec:04_results_baselines}, \cref{sec:results}].
    \item[$\circ$] A post-process relighting formulation that fully decouples relighting from avatar
    modeling via LoRA fine-tuning of a pre-trained video diffusion model, without any explicit
    physically-based light transport modeling [\cref{sec:method}].
    \item[$\circ$] Ablations on the approach, characterizing the importance of the diffusion prior and
    key training design choices [\cref{sec:studies}].
\end{itemize}

%% file: Main_Sections/20_related_work.tex
\section{Related Work}

\subsection{Full-body Human Avatars}

\noindent\textbf{Non-relightable Avatars} assume a flat-lit illumination~\cite{alldieck2018detailed,
habermann2021ddc, pang2024ash, junkawitsch2025eva, shao2025degas, shen2023xavatar, zheng2023AvatarReX,
habermann2023hdhumans, peng2021animatableneuralradiance, li2024animatablegaussians}, simplifying the
problem of human reconstruction. Template meshes, textures, and skinning techniques~\cite{loper2015smpl,
SMPL-X:2019, alldieck2018detailed, kavan2007dqs} form the foundation for modeling and posing a virtual
character. With the advent of neural networks, subsequent works employ neural components to model
pose-dependent geometry~\cite{habermann2021ddc, shen2023xavatar, peng2021animatableneuralradiance} and
enhance appearance modeling~\cite{shen2023xavatar, zheng2023AvatarReX}. The integration of neural
rendering techniques~\cite{nerf, kerbl20233d} in particular has led to significantly improved visual
quality~\cite{pang2024ash, junkawitsch2025eva, shao2025degas, li2024animatablegaussians,
habermann2023hdhumans, peng2021animatableneuralradiance}. Recent methods produce high-quality
renderings~\cite{pang2024ash, junkawitsch2025eva, shao2025degas}, model complex cloth
dynamics~\cite{habermann2021ddc, habermann2023hdhumans, pang2024ash, junkawitsch2025eva}, and support
explicit control over facial expressions~\cite{shen2023xavatar, junkawitsch2025eva, shao2025degas,
zheng2023AvatarReX}.

\noindent\textbf{Relightable Avatars} aim to construct a full intrinsic decomposition of a human, modeling
materials, environment illumination, and light transport~\cite{xu2023relightableanimatableneuralavatar,
lin2023relightableanimatableneuralavatars, chen2024meshavatarlearninghighqualitytriangular,
Wang2024IntrinsicAvatar, WangARXIV2025fullbodyrgca, iqbal2022ranarelightablearticulatedneural,
Luvizon2024relightneuralactor, Jiang2025DNF, singh2025rhc}. In practice, this typically involves
estimating albedo, roughness, and metalness, used alongside HDR environment maps or light volumes to
perform physically-based rendering with an analytic BRDF. However, recovering accurate, pose-dependent
intrinsics from multi-view observations is highly ill-posed: material-illumination ambiguities and
sensitivity to geometry errors make optimization difficult to constrain, causing most PBR-based methods
to fall short of the visual quality achieved by flat-lit
avatars~\cite{iqbal2022ranarelightablearticulatedneural, Luvizon2024relightneuralactor,
Wang2024IntrinsicAvatar, chen2024meshavatarlearninghighqualitytriangular,
lin2023relightableanimatableneuralavatars, xu2023relightableanimatableneuralavatar, Jiang2025DNF}. Singh
et al.~\cite{singh2025rhc} propose a neural light transport model, but require multi-view input during
inference. Wang et al.~\cite{WangARXIV2025fullbodyrgca} achieve high-quality PBR-based relighting, but
depend on highly accurate mesh tracking and carefully tuned optimization constraints, and do not
demonstrate strong facial expressions.

In contrast to existing relightable avatars, \ours~avoids physically-based components entirely, instead
learning an image-space relighting function that translates albedo-like renderings to a target
illumination. The function is trained directly on captured multi-illumination data, requiring neither
accurate geometry tracking nor additional regularization losses. Moreover, \ours~inherits all
capabilities of the underlying albedo avatar~\cite{junkawitsch2025eva}, including dynamic clothing and
facial expression control.

\subsection{Diffusion-based Relighting}

In recent years, generative diffusion models~\cite{podell2023sdxlimprovinglatentdiffusion,
peebles2023scalablediffusionmodelstransformers} have been leveraged by researchers for learning
relighting across a wide range of domains, including portrait~\cite{mei2025lux, wang2025comprehensive,
He2024DifFRelight}, full-body~\cite{wang2025comprehensive}, and general scenes and
objects~\cite{kocsis2024intrinsicimagediffusionindoor, zeng2024rgb, fang2025vrgbxvideoeditingaccurate,
liang2025DiffusionRenderer, he2025unirelight, pirier2024diffusionradiancefiledrelighting,
jin2024neural_gaffer, zeng2024dilightnet, tang2025rogr, zhang2024relitlrmgenerativerelightableradiance}.
Despite achieving photorealistic results, these methods typically do not enforce view or temporal
consistency. Wang et al.~\cite{wang2025comprehensive} produce temporally consistent video relighting but
do not demonstrate view consistency. Tang et al.~\cite{tang2025rogr} achieve multi-view consistency but
target only static objects. Most relevant to our work, He et al.~\cite{He2024DifFRelight} demonstrate
view-consistent relighting of person-specific flat-lit portraits under novel views and illuminations, but
are limited to faces and OLAT illuminations.

Inspired by these works, \ours~adapts video diffusion to achieve person-specific, view- and temporally
consistent full-body relighting via lightweight LoRA fine-tuning on flat-lit to HDR frame pairs,
leveraging the strong prior of a pre-trained relighting model~\cite{liang2025DiffusionRenderer}.

%% file: Main_Sections/30_method.tex
\section{Method}
\label{sec:method}
\input{Main_Figures/method_overview}

\ours~is a subject-specific video generative relighting framework designed to achieve temporally and spatially consistent, drivable, and expressive avatar relighting. Specifically, given driving dynamics --- including skeleton poses $\boldsymbol{\theta}^{\triangleright}$, FLAME expressions $\boldsymbol{\psi}^{\triangleright}$, HDR environment maps $\boldsymbol{l}^{\triangleright}$, and virtual viewpoints $v^{\triangleright}$ --- \ours~is trained on subject-specific real captures to synthesize photorealistic and temporally coherent relit videos. To this end, we first collect a high-quality dataset containing four dynamic subjects with strong expressions as supervision signals (\cref{sec:30_data_capture}). We then formulate relighting as an independent post-processing stage where the albedo-like avatar video $\mathcal{I}^{\triangleright}_{\text{WL}}$ (\cref{sec:30_white_light_avatar}), obtained from the driven poses and camera parameters, is mapped to relit video via generative relighting priors (\cref{sec:30_diffusion_relighting}) fine-tuned on the collected real captures. A full overview of the method is illustrated in~\cref{fig:method_overview}.

\subsection{Data Capture}
\label{sec:30_data_capture}

\subsubsection{Capture Protocol.}
To supervise the generative relighting framework and facilitate translation from flat-lit illumination $\mathcal{I}^{\triangleright}_{\text{WL}}$ to arbitrary illumination, we carefully collect subject-specific paired datasets of flat-lit and randomly sampled illuminations. Specifically, following Singh et al.~\cite{singh2025rhc} we collect
\begin{align}
&\mathcal{C}=\{(\mathcal{P}_{t} =(\mathcal{I}_{\text{WL}, t}; \mathcal{I}_{\text{RL},t}), v_{k}, \boldsymbol{l}_{t})\}\\
&0<t\leq N_{\text{pairs}}; 0<k\leq N_{\text{views}};\mathcal{I}_{\{\text{WL}, \text{RL}\}, t} \in  \mathbb{R}^{H\times W\times3}, v_{t} \in \mathcal{V}
\end{align}
where the full capture $\mathcal{C}$ consists of videos of sequential flat-lit to relit frame pairs $\mathcal{P}_{t}$, alongside corresponding calibrated camera parameters $v_{k}$ per view and HDR maps $\boldsymbol{l}_{t}$ per frame. $\mathcal{P}^{\triangleright}$ is obtained by interleaving white-light illuminated frames $\mathcal{I}_{\text{WL}}$ with frames $\mathcal{I}_{\text{RL}}^{\triangleright}$ illuminated under HDR maps $\boldsymbol{l}_{t}$. As the original protocol by Singh et al. does not include expressions, we follow the strategy of Junkawitsch et al.~\cite{junkawitsch2025eva} and instruct the subject to perform a list of pre-defined actions jointly with random facial expressions. We capture a total of $4$ subjects exhibiting a large range of clothing detail and dynamics, and collect $14400$ frame-pairs for each. Visuals of the data are provided in~\cref{fig:method_overview}. For more details and visuals regarding the data, we refer to the supplementary.\vspace{-5pt}

\subsubsection{Capture Setup.}
All our multi-illumination captures are recorded in a spherical dome equipped with $40$ RED Komodo 6K cameras and $331$ individually controllable LEDs capable of emitting red, green, blue, amber, and white light (RGBAW). The cameras and lights are arranged $360^{\circ}$ around the subject, enabling the capture of synchronized multi-illumination sequences at $30$ FPS at $4$K resolution. For Subject 1, all cameras observe the full-body human, while for Subjects 2, 3, and 4, three front-facing cameras are additionally focused on the face. For an illustration of our capture setup, please view the supplementary.

\subsubsection{Data Preprocessing.}

After camera calibration and undistortion, we extract frame-wise body poses $\boldsymbol{\theta}_{\text{Full}}^{\triangleright} = \boldsymbol{\theta}_{\text{WL}}^{\triangleright} \cup \boldsymbol{\theta}_{\text{RL}}^{\triangleright}$ and initial segmentation masks $\mathcal{S}_\text{Full}^{\triangleright} = \mathcal{S}_{\text{WL}}^{\triangleright} \cup \mathcal{S}_{\text{RL}}^{\triangleright}$ using Captury~\cite{captury} and the Sapiens foundation model~\cite{khirodkar2024sapiensfoundationhumanvision}. While the segmentation method works well for flat-lit frames, we observe it consistently produces noisy results for underexposed HDR frames. As such, we additionally refine $\mathcal{S}_{\text{Full}}$ by filtering out high-frequency flickers in the temporal dimension. From the flat-lit RGB frames $\mathcal{I}_{\text{WL}}^{\triangleright}$ and associated masks $\mathcal{S}_{\text{WL}}^{\triangleright}$ we also construct reference geometry $\textbf{V}_{\text{WL},\text{Ref.}}^{\triangleright}$ using the method of Wang et al.~\cite{wang2023neus2}. Moreover, we estimate a high-detail template mesh $\textbf{V}_{\text{Templ.}}$ of the subject in A-pose via a commercial scanner setup~\cite{treedys} and multi-view stereo reconstruction software~\cite{agisoft_metashape}. Finally, following Junkawitsch et al.~\cite{junkawitsch2025eva} we further refine the flat-lit body poses $\boldsymbol{\theta}_{WL}^{\triangleright}$ based on the skinned template mesh~\cite{kavan2007dqs} and reference geometry into $\hat{\boldsymbol{\theta}}_{WL}^{\triangleright}$, and regress FLAME expression parameters $\boldsymbol{\psi}_{WL}^{\triangleright} = (\theta_{jaw}, \gamma, \epsilon)$ from the flat-lit RGB frames using FAN~\cite{Bulat_2017_Fan} and Mediapipe~\cite{lugaresi2019mediapipeframeworkbuildingperception} landmarks.

\subsection{Albedo Driving Avatar}
\label{sec:30_white_light_avatar}

To close the domain gap between rendered flat-lit (albedo-like) images and real flat-lit images, and to enable expression control in the final relightable avatar, a suitable albedo-like avatar is required. To this end, \ours~adopts \textit{Expressive Virtual Avatars (EVA)}~\cite{junkawitsch2025eva}. As seen in~\cref{fig:method_overview}, \textit{EVA} is composed of an expressive animatable template geometry layer~\cite{habermann2021ddc, FLAME} which models coarse geometry, and a disentangled UV Gaussian appearance layer~\cite{kerbl20233d, pang2024ash}. The latter consists of two U-Nets (face and body) trained to translate motion-aware textures (a short sequence of normal and position UV textures) to predict Gaussian attributes in UV space, which are anchored to the deformable template and splatted.

Following the original work, we train \textit{EVA} using the flat-lit frames of our light stage capture. For efficient training, we downsample images to a resolution of $540\times1024$ and incorporate high-resolution crops to further enhance appearance reconstruction. We observe that slightly overestimated segmentation masks (\cref{sec:30_data_capture}) lead to the creation of background Gaussian and floater artifacts. As such, we slightly modify the training scheme to optimize using background-composited images starting after $40000$ training steps:
\begin{equation}
    \hat{\mathcal{I}}_{\text{Pred}} = \mathcal{I}_{\text{Pred}} \cdot \mathcal{I}_{\alpha} + \mathcal{I}_{\text{BG}} \cdot ( 1-\mathcal{I}_{\alpha})
\end{equation}
Here, $\mathcal{I}_{\text{Pred}}, \mathcal{I}_{\alpha}$ denote the originally predicted RGB and alpha images, and $\mathcal{I}_{\text{BG}}$ is the ground truth background with the overestimated mask applied. In the following optimization, $\mathcal{I}_{\text{Pred}}$ is replaced by $\hat{\mathcal{I}}_{\text{Pred}}$.

\subsection{Diffusion Relighting}
\label{sec:30_diffusion_relighting}

In this subsection, we obtain a subject-specific generative relighting framework able to translate flat-lit dynamic avatar renders $\mathcal{I}^{\triangleright}_{\text{WL}}$ to arbitrary illuminations ${l}^{\triangleright}$. As recent works in video diffusion relighting demonstrate remarkable visual quality alongside temporal coherence~\cite{he2025unirelight, liang2025DiffusionRenderer, fang2025vrgbxvideoeditingaccurate, wang2025comprehensive, mei2025lux}, we propose to adapt the strong pre-trained prior of a general object/scene generative video relighting model to our task. Specifically, \ours~adopts Cosmos DiffusionRenderer (DRCosmos)~\cite{liang2025DiffusionRenderer}, a video diffusion model trained on a large-scale dataset, which has demonstrated remarkable relighting performance on object- and scene-specific benchmarks. However, it is not directly compatible with the full-body avatar relighting task due to the distribution gap between avatar-specific captures and its original training data. To address this, we fine-tune DiffusionRenderer on our captured real data (\cref{sec:30_data_capture}), thereby adapting it to the full-body avatar relighting setting.

DRCosmos consists of two DiT-based modules~\cite{peebles2023scalablediffusionmodelstransformers}: an inverse module mapping RGB video to G-Buffers (albedo, metalness, roughness, and normal) and a forward renderer mapping G-Buffers to RGB video.
In our framework, we discard the inverse module and only fine-tune the LoRA weights of the forward renderer, where we treat the flat-lit avatar, obtained in \cref{sec:30_white_light_avatar}, as an approximation of albedo and input to the forward renderer (albedo avatar).
Consequently, we retain only the base color channel and discard the remaining G-buffers by setting them to $0$.
As illustrated in \cref{fig:method_overview}, input attributes ($\mathcal{I}_{\text{WL}}$, $\boldsymbol{l}$) are encoded to their corresponding latent representations using the DRCosmos VAE encoder $\mathcal{E}$ and concatenated with a sequence of random noisy latents $\mathbf{z}_{\tau}$, as the input to the DiT.
During training, we randomly sample instances from the captured data pairs $(\mathcal{I}_{\text{WL, i}}; \mathcal{I}_{\text{RL, i}}; l)$, where clean target frames $\mathcal{I}_{\text{RL}}$ are corrupted with noise $\tau$ to reconstruct noisy latents $\mathbf{z}_{\tau}$. Following the formulation of DRCosmos, the training loss for our generative relighting module is defined as:
\begin{equation}
    \mathcal{L}(\Delta\theta) = \Vert \mathbf{f}_{ \Delta\theta} (\mathbf{z}_{\tau}, \mathbf{g}, \mathbf{c_{\text{env}}},\tau) - \mathbf{z}_{0} \Vert_{2}^{2},
\end{equation}
where $\mathbf{f}_{\Delta\theta}$ is the denoising function employed by the model with added LoRA parameters $\Delta\theta$, and $\textbf{g}, \textbf{c}_{\text{env}},\mathbf{z}_{0}$ correspond to the encoded latents of $\mathcal{I}_{\text{WL}}$, $\tilde{\boldsymbol{l}}_{i}$, and $\mathcal{I}_{\text{RL}}$. $\tilde{\boldsymbol{l}}$ denotes the world-space HDR maps $\boldsymbol{l}$ projected to views $v$ and tone-mapped to low dynamic range using Reinhard tone-mapping.

As the model was originally trained on primarily single-image and $57$-frame chunk batches at a resolution of $1280\times704$, we downsample our images to $371\times704$ and pad with $0$ on both sides to match the default configuration as closely as possible. We empirically observe that even when the model is only fine-tuned on single-image frame pairs, it preserves temporal consistency in views and illumination across $57$-frame chunks, as demonstrated in the original work. While some illumination changes are present across inferences, the variation is low enough for \ours~to render perceptually convincing long-form videos via linear blending using an overlap of $32$ frames.

%% file: Main_Figures/method_overview.tex
\begin{figure}[t]
    \includegraphics[width=\textwidth]{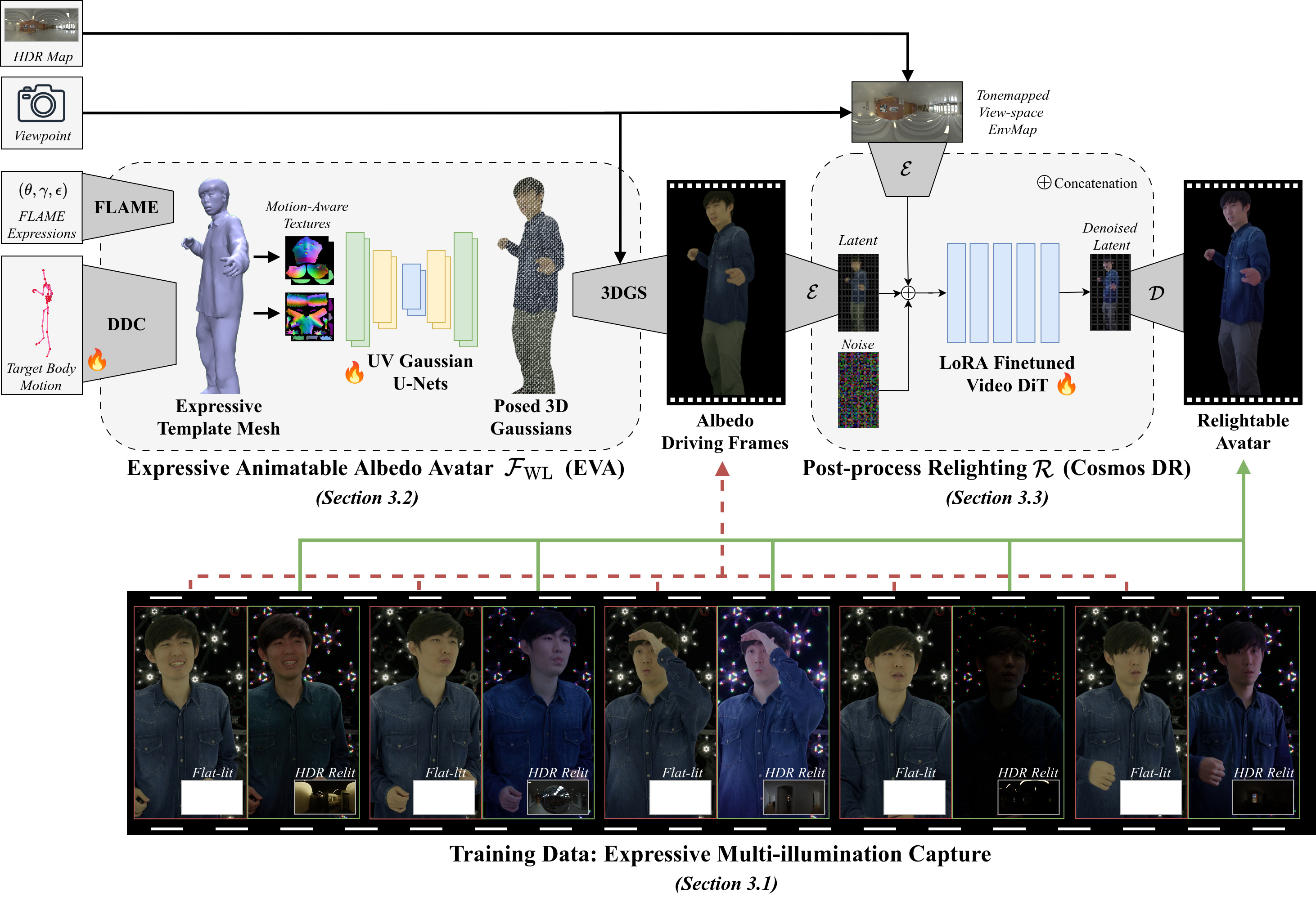}
    \vspace{-10pt}
    \caption{\ours{} overview. Given a calibrated sequence of flat-lit and HDR-illuminated multi-view frame
  pairs (\cref{sec:30_data_capture}), \ours{} trains two independent components. An expressive, controllable
  albedo avatar $\mathcal{F}_{\text{WL}}$ (\textit{EVA}~\cite{junkawitsch2025eva}) is trained on the
  flat-lit frames to render albedo-like images for arbitrary pose, expression, and viewpoint. In parallel,
  a video diffusion relighting model~\cite{liang2025DiffusionRenderer} is fine-tuned via LoRA on the
  flat-lit to relit frame pairs, learning a person-specific relighting function $\mathcal{R}$ that
  translates flat-lit images to a target illumination. At inference, $\mathcal{R}$ is applied to renderings
  from $\mathcal{F}_{\text{WL}}$, yielding the relightable avatar $\mathcal{F}_{\text{RL}}$.}
    \label{fig:method_overview}
    \vspace{-10pt}
\end{figure}

%% file: Main_Sections/40_results.tex
\section{Results}

We present quantitative and qualitative evaluations of \ours. Implementation and training details are given in~\cref{sec:implement}, qualitative results in~\cref{sec:results}, baseline comparisons in~\cref{sec:04_results_baselines}, and studies on the diffusion relighting model in~\cref{sec:studies}.

\subsection{Implementation and Training Details}
\label{sec:implement}

We train a \ours~avatar for each captured subject. For all evaluations, we withhold $\sim\!10\%$ of lighting conditions (uniformly sampled) and motion frames (i.e.\ frame pairs with index $\geq 12250$), as well as $4$ full-body cameras, from training.
All modules are implemented in PyTorch. The deformable body model~\cite{habermann2021ddc} for \textit{EVA} takes approximately $4$ days to train on $4$ NVIDIA A40 GPUs, while \textit{EVA}'s Gaussian appearance layer~\cite{junkawitsch2025eva} takes approximately $3$ days on $1$ NVIDIA A40 GPU. LoRA fine-tuning the diffusion relighting model~\cite{liang2025DiffusionRenderer} to convergence takes approximately $2$ days on $1$ NVIDIA H100 GPU with a batch size of $5$.

\subsection{Qualitative Results}
\label{sec:results}

We illustrate view, illumination, and motion consistency across a 57-frame rendered video in~\cref{fig:all_olat}, showing $2$ of our $4$ subjects rendered under close-up and far-field views, relit under a single directional area light. HDR environment map-based relighting across a wide range of motions, expressions, views, and illuminations for all $4$ subjects is shown in~\cref{fig:qual_hdri}. \cref{fig:view_consistency} provides an illustration of consistency between separate inference runs. Each experiment figure (\cref{fig:baselines_01,fig:ablation_enhance,fig:ablation_prior_qual,fig:ablation_masks}) additionally provides far-field and close-up views to facilitate inspection of view consistency. For additional qualitative results on cross-inference view, illumination, and motion consistency, as well as long-form video with linear blending, we refer to the supplementary material.

We observe that the diffusion model retains strong consistency across view, illumination, and motion within a 57-frame chunk. Across separate inferences, view consistency remains strong, while illumination and motion exhibit more noticeable variance. We conjecture that this variance is primarily caused by subject drift between flat-lit and relit frames in the training data. Nevertheless, the variance is low enough to be convincingly suppressed via simple linear blending. Overall, \ours{} achieves photorealistic relightable avatar rendering.

\subsection{Baselines}
\label{sec:04_results_baselines}

\input{Main_Figures/qual_baselines_01}
\input{Main_Tables/quant_baseline_comp}

\subsubsection{Setup} To our knowledge, no existing public method can directly consume our data format (expressive multi-view multi-illumination captures). We therefore manually construct the following baselines: (1) We adapt MeshAvatar~\cite{chen2024meshavatarlearninghighqualitytriangular} as a PBR-based baseline by excluding the lighting condition from optimization and setting it explicitly to our ground truth. The required SMPL-H poses are optimized using EasyMocap~\cite{easymocap} against 3D OpenPose keypoints~\cite{cao2019openpose} tracked against our skeleton. (2) To evaluate the contribution of \textit{EVA} as the driving avatar, we train a MeshAvatar instance only on flat-lit frames, which then replaces \textit{EVA} as the albedo driving avatar in \ours. (3) Finally, we combine the white-light \textit{EVA} avatar with IC-Light~\cite{zhang2025scaling} as a zero-shot relighting baseline, providing the relit light stage background as conditioning for background harmonization. We acknowledge that stronger baselines exist~\cite{wang2025comprehensive, kim2024switchlight, WangARXIV2025fullbodyrgca}, but the absence of publicly available code precludes direct comparison. We test all methods under two settings: (a) novel view/motion under a light seen during training and (b) novel view/motion/light. For each, we provide PSNR, SSIM, and LPIPS against masked image regions of interest.

\subsubsection{Results} Quantitative baseline comparisons under novel view/motion and novel view/motion/light settings are provided in~\cref{tab:quant_baseline_compare}, while qualitative results under the novel view/motion/light setting are shown in~\cref{fig:baselines_01}. For additional illustrations, we refer to the supplementary. \ours{} produces significantly more realistic results than all baselines, which is supported by the quantitative evaluation. Both MeshAvatar-based baselines (1 + 2) fail to accurately reconstruct loose clothing and expressive faces. The PBR baseline (1) produces flat appearance and fails to model complex illumination effects---e.g.\ Fresnel reflectance and specular highlights---to the same degree as the diffusion-based relighting step. The PBR optimization is also significantly harder to constrain, causing geometry artifacts in areas not consistently well-lit during HDR capture. The diffusion relighting baseline (2) achieves substantially improved realism, even approaching \ours~quantitatively and qualitatively. Interestingly, despite being trained only as a relighting function, we observe the diffusion model starts to act as an enhancement model, introducing finer detail in face and hand regions. Finally, the zero-shot IC-Light baseline (3) largely preserves \textit{EVA} rendering quality but consistently produces overexposed results and does not produce view-consistent renderings.

\subsection{Ablation Studies}
\label{sec:studies}

We conduct ablation studies on the diffusion relighting model, investigating: (A) the importance of the pre-trained prior, (B) training the relighting step to additionally perform enhancement of \textit{EVA} renderings, and (C) training the diffusion model with and without a masked background.

\subsubsection{A) The importance of the prior} is evaluated by comparing four model variants: (1) zero-shot with no fine-tuning, (2) full fine-tuning of all network parameters from the pretrained checkpoint, (3) full fine-tuning from random weight initialization, and (4) LoRA fine-tuning (ours). For (2) and (3), we train on two NVIDIA H100 GPUs with a batch size of $2$ for $30$k iterations each; variant (4) is trained on one GPU with a batch size of $5$. Two of the four subjects are evaluated.

Qualitative and quantitative results are provided in~\cref{fig:ablation_prior_qual} and~\cref{tab:quant_ablation_dr_prior}. Both the LoRA variant (4) and full fine-tuning from the pretrained checkpoint (2) converge successfully, producing comparable results. Meanwhile, full fine-tuning from random initialization (3) fails to converge within the allocated training budget. Without fine-tuning (1), the model cannot relight realistically or consistently across inference runs, yet exhibits some understanding of light transport and remains temporally stable within a 57-frame chunk, as shown in the supplementary video.

\subsubsection{B) Training the relighting model for enhancement} is done by learning a direct \textit{EVA}-to-relit translation function, obtained by replacing the captured flat-lit frames with \textit{EVA} renderings during training. We evaluate each variant on the trained subject and its generalization to the remaining subjects, averaging across all $4$ subjects.
As shown in~\cref{fig:ablation_enhance}, the enhancement variant generally achieves higher visual quality by correcting \textit{EVA} artifacts. However, as \textit{EVA} renderings are less well-aligned to the relit frames than their flat-lit counterparts, the enhancement model also tends to introduce stronger spurious artifacts, most noticeably in the eyes since \textit{EVA} does not track gaze. As shown in~\cref{tab:quant_ablation_dr_coupling}, the enhancement model outperforms the relighting-only version on $3$ of the $4$ subjects, but produces more color shifts on subject $1$.

\subsubsection{C) The ablation on background masks} consists of training the diffusion model once on frame pairs directly taken from the capture (w/o masks), and once on frame pairs with the background elements masked out (w/ masks). Again, all four subjects are evaluated and averaged.

As shown in~\cref{fig:ablation_masks}, the masked variant (w/ masks) more accurately matches ground truth color when driven by \textit{EVA} renderings, supported by the quantitative results in~\cref{tab:quant_ablation_dr_masks}. The unmasked variant (w/o masks) still achieves reasonable perceptual relighting quality and robustly retains the black background of \textit{EVA} renders. Consequently, training without masks may be more suitable for scaling, as it requires no additional image preprocessing.

\input{Main_Figures/ablation_masks}
\input{Main_Tables/quant_ablation_dr_masks}

\input{Main_Figures/ablation_enhance}
\input{Main_Tables/quant_ablation_dr_coupling}

\input{Main_Figures/ablation_prior}
\input{Main_Tables/quant_ablation_dr_prior}

\input{Main_Figures/views}
\input{Main_Figures/qual_olats}

%% file: Main_Figures/qual_baselines_01.tex
\begin{figure}
    \centering
    \includegraphics[width=0.95\textwidth]{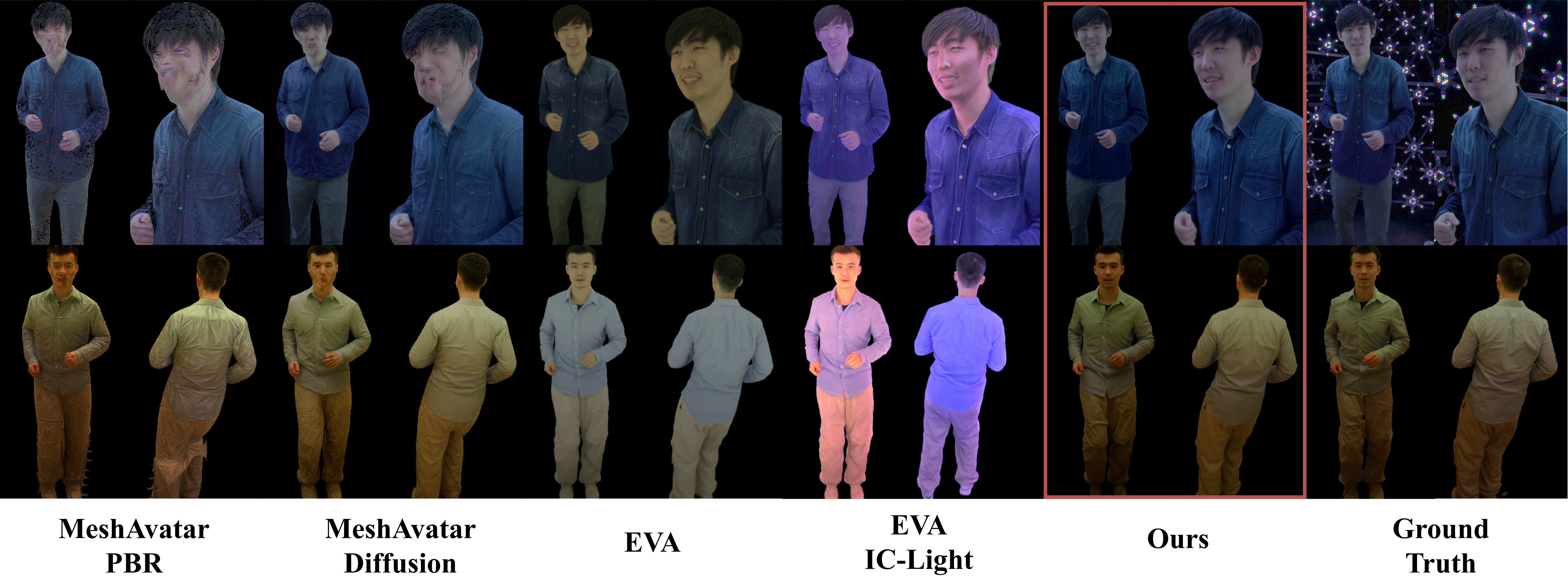}
    \caption{Qualitative comparison against our constructed baselines. The two MeshAvatar-based     baselines~\cite{chen2024meshavatarlearninghighqualitytriangular} are unable to reconstruct the face due to missing expression dependence. The PBR version of MeshAvatar, adapted to our setting, is able to learn intrinsics for relighting, but optimization is difficult to constrain. In comparison, training this version of MeshAvatar of flat-lit frames only, with video diffusion for relighting, achieves significantly improved results, demonstrating the efficacy of our approach. The zero-shot baseline consists of IC-Light~\cite{zhang2025scaling} with \textit{EVA}~\cite{junkawitsch2025eva} and light stage background as conditioning, which is unable to relight realistically and consistently.}
    \label{fig:baselines_01}
    \vspace{-15pt}
\end{figure}

%% file: Main_Tables/quant_baseline_comp.tex
\begin{table}[t]
    \centering
    \small
    \setlength{\tabcolsep}{3pt} 
    \caption{Comparison results against the constructed baselines: MeshAvatar~\cite{chen2024meshavatarlearninghighqualitytriangular} adapted to our calibrated illumination setting [MA/PBR], a post-process diffusion relighting version of MeshAvatar [MA/DR] and the zero-shot relighting method IC-Light~\cite{zhang2025scaling} applied to our driving EVA renderings [EVA/IC]. The evaluation is averaged across all 4 subjects.}
    \begin{tabular}{lcccccc}
        \toprule
        
        & \multicolumn{3}{c}{\textbf{Novel View/Motion}} 
        & \multicolumn{3}{c}{\textbf{Novel View/Motion/Light}} \\
        
        \cmidrule(lr){2-4} 
        \cmidrule(lr){5-7}
        
        & PSNR $\uparrow$ & LPIPS $\downarrow$ & SSIM $\uparrow$
        & PSNR $\uparrow$ & LPIPS $\downarrow$ & SSIM $\uparrow$ \\
        
        \midrule
        MA/PBR~\cite{chen2024meshavatarlearninghighqualitytriangular}
        & 22.61 & 0.095 & 0.899
        & 22.38 & 0.095 & 0.899 \\

        MA/DR~\cite{chen2024meshavatarlearninghighqualitytriangular, liang2025DiffusionRenderer} 
        & \underline{26.07} & \underline{0.080} & \underline{0.935}  
        & \underline{25.73} & \underline{0.080} & \underline{0.933}  \\

        EVA/IC
        & 10.26 & 0.108 &  0.853
        &  10.41 & 0.106 & 0.855  \\
        
        \midrule
        \textbf{Ours}    
        & \textbf{26.62} & \textbf{0.077} &  \textbf{0.937}
        & \textbf{26.30} & \textbf{0.077} &  \textbf{0.936} \\
        
        \bottomrule
    \end{tabular}
    \vspace{4pt}
    
    \label{tab:quant_baseline_compare}

\end{table}

%% file: Main_Figures/ablation_masks.tex
\begin{figure}[t]
    \centering
    \includegraphics[width=\textwidth]{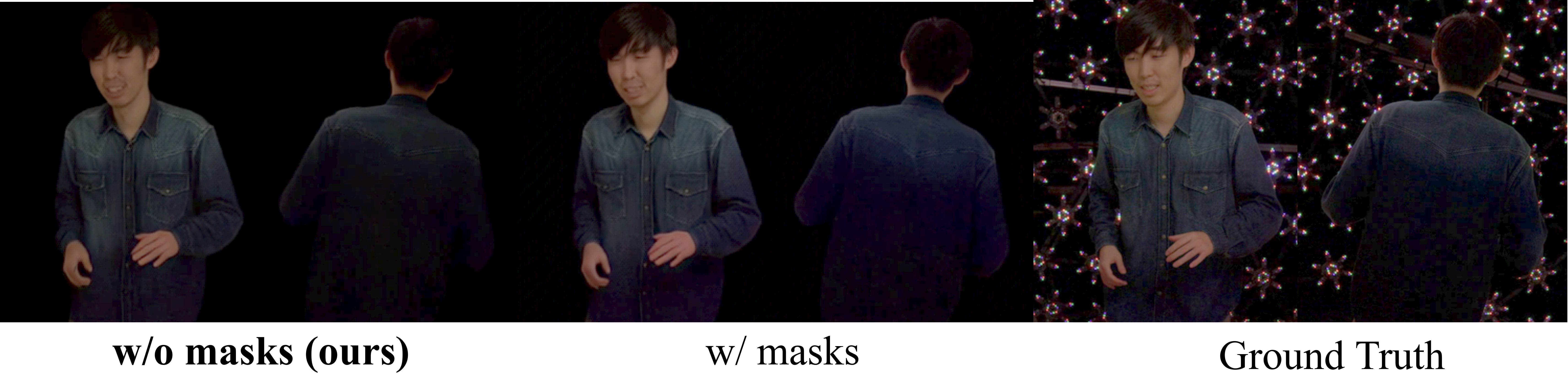}
    \caption{Qualitative results for the experiment on removing background elements. While the model achieves results closer to ground truth by removing the background, perceptually realistic relighting can be achieved only through fine-tuning on the unmasked frame pairs.}
    \label{fig:ablation_masks}
    \vspace{-5pt}
\end{figure}

%% file: Main_Tables/quant_ablation_dr_masks.tex
\begin{table}[t]
    \centering
    \small
    \setlength{\tabcolsep}{3pt} 
        \caption{Quantitative comparison on the effect of removing background elements (w/ BG masks) vs retaining (w/o BG masks) them in training frame pairs. While w/ BG consistently achieves results closer to ground truth, w/o still achieves perceptually convincing renders and crucially, can be run directly on the captures.}.
    \begin{tabular}{lcccc}
        \toprule
         $\mathcal{R}$ target task & PSNR $\uparrow$ & LPIPS  $\downarrow$ & SSIM $\uparrow$ & Preprocess Time $\downarrow$ \\
        \midrule
        \textbf{w/o BG masks  (Ours)}
        & 26.30 & 0.077 & 0.936  & \textbf{0s}  \\
        w/ BG masks
        & \textbf{26.82} & \textbf{0.073} & \textbf{0.939} & $16$ Days \\
        \bottomrule
    \end{tabular}

    \label{tab:quant_ablation_dr_masks}
\end{table}

%% file: Main_Figures/ablation_enhance.tex
\begin{figure}[t]
    \centering
    \includegraphics[width=\textwidth]{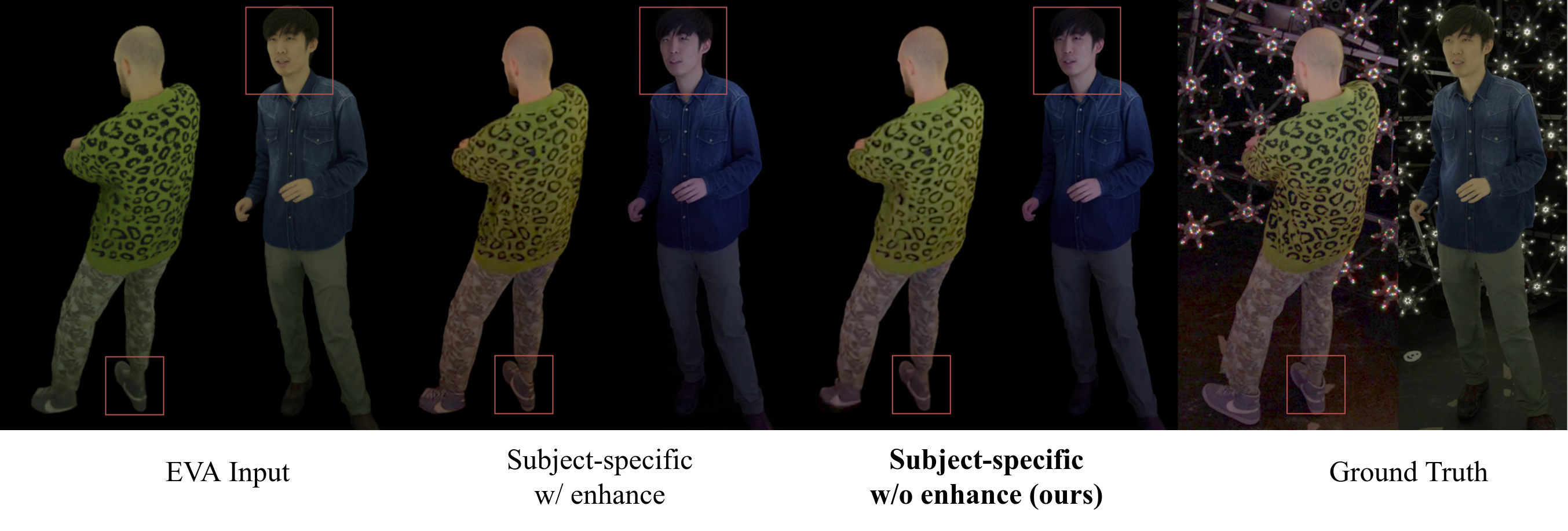}
    \vspace{-20pt}
    \caption{Qualitative results for training the video diffusion model on \textit{EVA} to relit (enhancement) vs. real flat-lit to relit (relight only) frame pairs. While the enhancement model can fix artifacts in \textit{EVA} (blurry shoes) it also tends to hallucinate more.}\vspace{-5pt}
    \label{fig:ablation_enhance}
\end{figure}

%% file: Main_Tables/quant_ablation_dr_coupling.tex
\begin{table}[h]
    \centering
    \small
    \setlength{\tabcolsep}{3pt} 
        \caption{Quantitative comparison on utilizing the video diffusion model for enhancement. The enhancement model is able to correct \textit{EVA} artifacts, resulting in lower LPIPS across subjects. However, it is less robust at relighting.}.
    \begin{tabular}{lcccc}
        \toprule
        
         Target task & PSNR $\uparrow$ & LPIPS  $\downarrow$ & SSIM $\uparrow$ & \textit{EVA} specific?\\
        
        \midrule
        \textbf{Relight only (Ours)}
        & \textbf{26.3} & 0.077 & \textbf{0.936} & \textbf{No}\\ 
        Relight + Enhance
        & 26.12 & \textbf{0.072} & 0.938  &  {Yes}\\
        \bottomrule
    \end{tabular}
    \vspace{4pt}

    \label{tab:quant_ablation_dr_coupling}
\end{table}

%% file: Main_Figures/ablation_prior.tex
\begin{figure}[t]
    \centering
    \includegraphics[width=\textwidth]{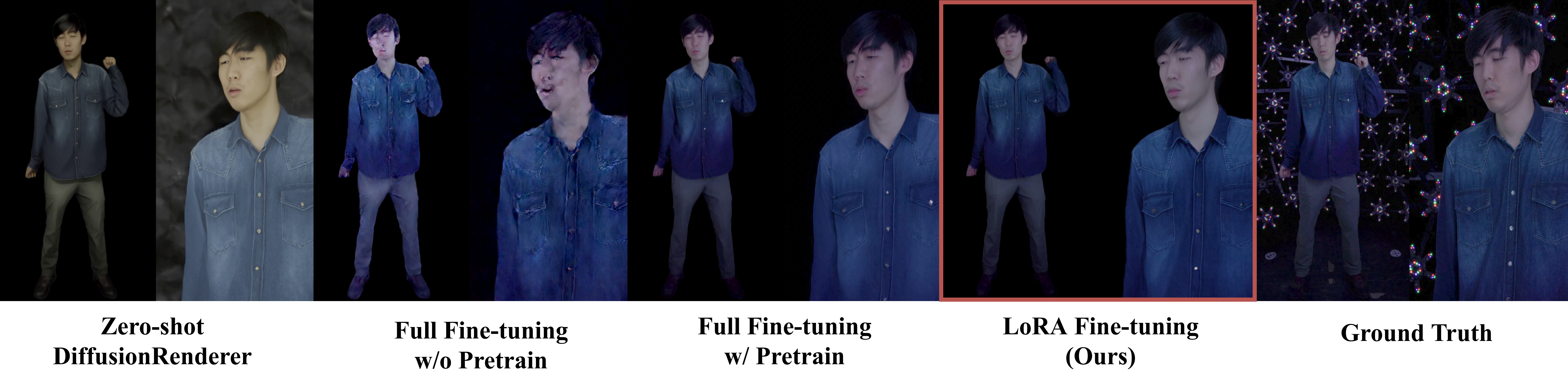}
    \caption{Qualitative results for different model training strategies. Due to the domain gap, without fine-tuning the diffusion model is unable to perform realistic relighting on the full-body human. Both LoRA and full model fine-tuning from the pretrained checkpoint allow the model to learn convincing relighting, while training from random initialization fails to properly converge.}\vspace{-15pt}
    \label{fig:ablation_prior_qual}
\end{figure}

%% file: Main_Tables/quant_ablation_dr_prior.tex
\begin{table}[t]
    \centering
    \small
    
    \setlength{\tabcolsep}{3pt} 
        \caption{Quantitative comparison on the importance of the pre-trained prior. Both without fine-tuning and without the prior, the model is unable to realistically relight the person. LoRA-based fine-tuning achieves results comparable to fine-tuning the full model, while requiring far fewer resources.}.
    \begin{tabular}{lcccc}
        \toprule
         $\mathcal{R}$ training strategy& PSNR $\uparrow$ & LPIPS  $\downarrow$ & SSIM $\uparrow$ & GPU Hours $\downarrow$\\
        
        \midrule
        \textbf{LoRA adapted DR (Ours)}
        &\underline{27.62} & \textbf{0.079} & \underline{0.945}   & $\boldsymbol{\sim} \textbf{48h}$\\
        (a) w/o finetuning
        & 19.18 & 0.120 & 0.889 & 0h \\
        (b) with full finetuning
        & \textbf{27.88} & \textbf{0.079} & \textbf{0.947} & $\sim 192\text{h}$\\
        (c) w/o prior 
        & 21.27 & 0.096 & 0.91 & $\sim 192\text{h}$\\
        \bottomrule
    \end{tabular}
    \vspace{4pt}

    \label{tab:quant_ablation_dr_prior}
\end{table}

%% file: Main_Figures/views.tex
\begin{figure}
    \centering
    \includegraphics[width=\textwidth]{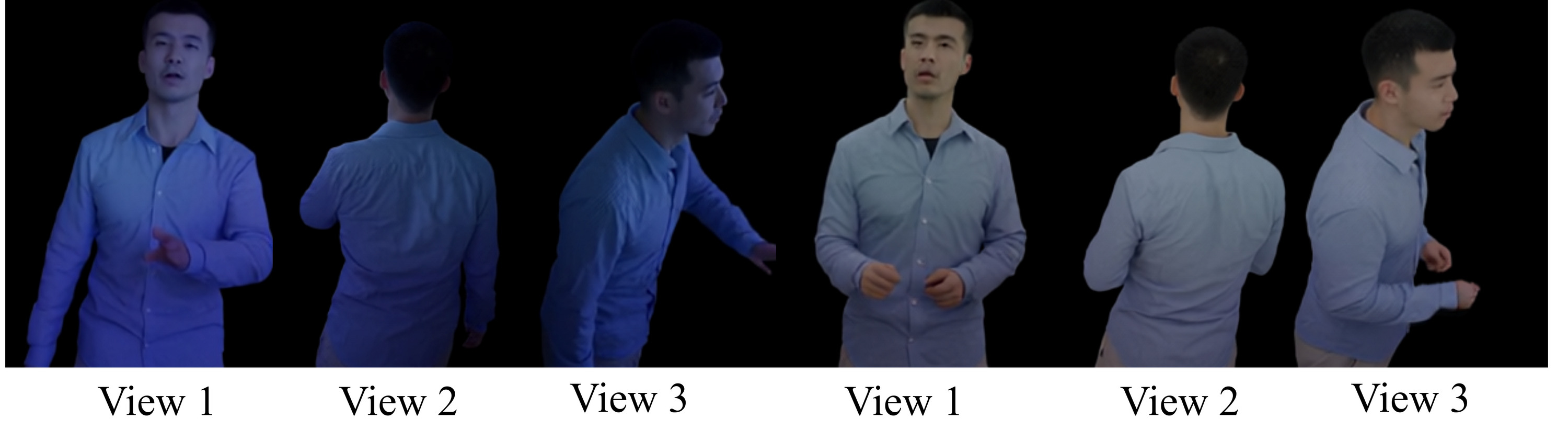}
    \vspace{-10pt}
    \caption{Qualitative demonstration of consistency across inference runs.}\vspace{-15pt}
    \label{fig:view_consistency}
\end{figure}

%% file: Main_Figures/qual_olats.tex
\begin{figure}[t]
    \centering
    \includegraphics[width=\textwidth]{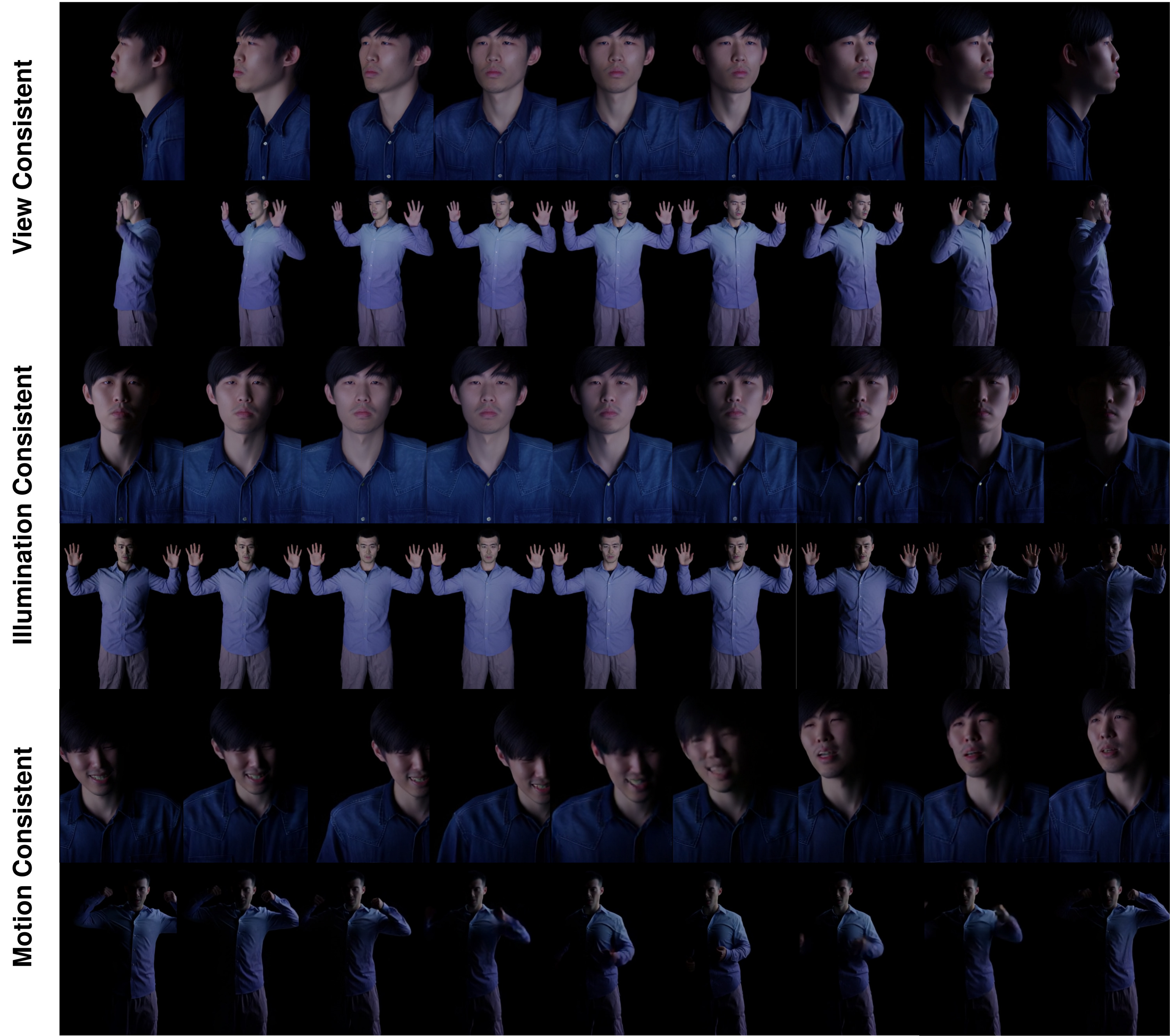}
    \caption{Qualitative results for consistent video diffusion relighting. Row 1: Results for a $\SI{180}{\degree}$ camera rotation under a fixed OLAT light, demonstrating view consistency. Row 2: Results for a $\SI{180}{\degree}$ OLAT light rotation under a fixed front-facing view, showing illumination consistency. Row 3: Results for a $57$-frame sequence under a fixed view and OLAT light, illustrating motion consistency. Fine-tuning on frame pairs is sufficient for the video diffusion model to retain consistency across all dimensions.}
    \label{fig:all_olat}
    \vspace{10pt}

    \includegraphics[width=\textwidth]{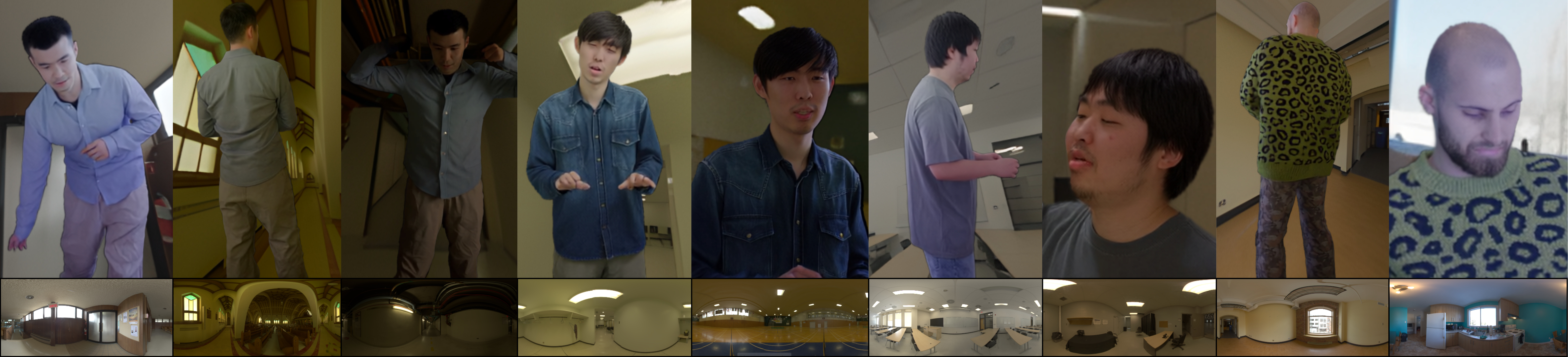}
    \caption{Results for direct HDRI relighting under diverse expressions and motions. \ours{} enables rendering of a person under controllable motion, expression, view, and illumination.}
    \label{fig:qual_hdri}
\end{figure}

%% file: Main_Sections/45_Limitations.tex
\section{Limitations and Future Work}
\label{sec:045_limitations}
\vspace{-10pt}While the video model preserves consistency well within a $57$-frame chunk, small shifts between separate inference passes remain present. As demonstrated in the supplementary video, linear blending can mask this effect, but requires a large chunk overlap and does not fully eliminate it. The magnitude of subject motion between the flat-lit and relit frame appears to be the primary factor influencing the variance (please view the supplementary for a detailed discussion). Conditioning the video model to produce a consistent continuation given a small set of starting frames could provide a satisfying solution to this problem. Moreover, we observe that light effects not well represented in our data --- such as far-away occlusion shadows --- are not learned. Crucially, although the proposed method is well-suited for offline applications (e.g., movie production), it does not support real-time rendering, which is essential for applications such as telepresence or VR/AR. The critical bottleneck is the video diffusion relighting model, which runs at ${\sim}0.5$ FPS under $15$ diffusion steps. Techniques such as deterministic one-step fine-tuning~\cite{liang2025DiffusionRenderer} and emerging methods for single-step relighting via flow matching~\cite{zhang2026renderflowsinglestepneuralrendering} hold promise to resolve this issue.
We observe that the fine-tuned diffusion relighting model requires sufficiently small shifts between flat-lit and relit frame pairs for stable inference. The issue could be addressed by adjusting the data collection protocol to record only slow movement, integrating emerging technology~\cite{digitalbipack2025yu}, and applying motion-based filtering during diffusion model optimization. Moreover, \ours~requires light stage capture to enable relighting, limiting accessibility. However, as demonstrated in~\cref{sec:04_results_baselines}, the video diffusion model requires only flat-lit to relit frame pairs and is therefore well-suited for training on large-scale data. Future work will investigate integrating existing static~\cite{teufelgera2025HumanOLAT, zhou2023relightable} and dynamic multi-illumination human datasets to test generalization.\vspace{-10pt}

%% file: Main_Sections/50_conclusion.tex
\section{Conclusion}
\vspace{-10pt}We presented \ours, a person-specific framework for photorealistic, relightable, expressive, and
animatable full-body human avatars. By treating relighting as an image-space post-process, \ours{}
sidesteps explicit intrinsic decomposition and instead learns a consistent relighting function that
translates flat-lit renderings from a driving avatar to a target HDR illumination, while fully inheriting
the expressive facial animation and dynamic clothing of the underlying avatar. We show that this can be
achieved solely through lightweight LoRA fine-tuning on person-specific flat-lit to relit frame pairs.

Leveraging the strong prior of a pre-trained general scene video diffusion relighting model, \ours{}
produces perceptually realistic, view- and temporally consistent relighting that outperforms a
physically-based rendering baseline with minimal training overhead. Looking ahead, we believe that
training on large-scale full-body multi-illumination data could enable a general relighting function
applicable to arbitrary humans, removing the need for person-specific fine-tuning.

%% file: Supp_Sections/00_overview.tex
\section{Overview}

This supplementary describes details for the data capture in~\cref{sec:supp_data}, gives additional quantitative and qualitative results in~\cref{sec:add_quant} and~\cref{sec:add_qual}, and provides an additional ablation on cross-subject generalization in~\cref{sec:generalization}. Moreover, we discuss the influence of misalignment between frame pairs in~\cref{sec:misalignment} and potential ethical concerns in~\cref{sec:ethics}.

%% file: Supp_Sections/01_data_capture.tex
\section{Data Capture Details}
\label{sec:supp_data}
Images visualizing our capture setup and collected data are provided in~\cref{fig:supp_datacap}. In increasing order of complexity, we collect the following subjects:
\begin{itemize}
    \item[$\bullet$] \textbf{Subject 1}: Male wearing a tight-fitting flat-color button-up shirt
    \item[$\bullet$] \textbf{Subject 2}: Male wearing a tight-fitting detailed button-up shirt
    \item[$\bullet$] \textbf{Subject 3}: Male wearing a semi-loose flat-color t-shirt
    \item[$\bullet$] \textbf{Subject 4}: Male wearing a highly loose and detailed pullover
\end{itemize}
All subjects feature mostly static hair and unoccluded faces. For subjects 2, 3, and 4, three front-view cameras of the total $40$ are focused on the upper body to capture additional detail. Moreover, we note that due to technical limitations with the capture setup, compared to subject 1, the overall intensity of lights is reduced for subjects 2, 3, and 4.

\input{Supp_Figures/data_capture}

%% file: Supp_Figures/data_capture.tex
\begin{figure}
    \centering
    \includegraphics[width=0.95\textwidth]{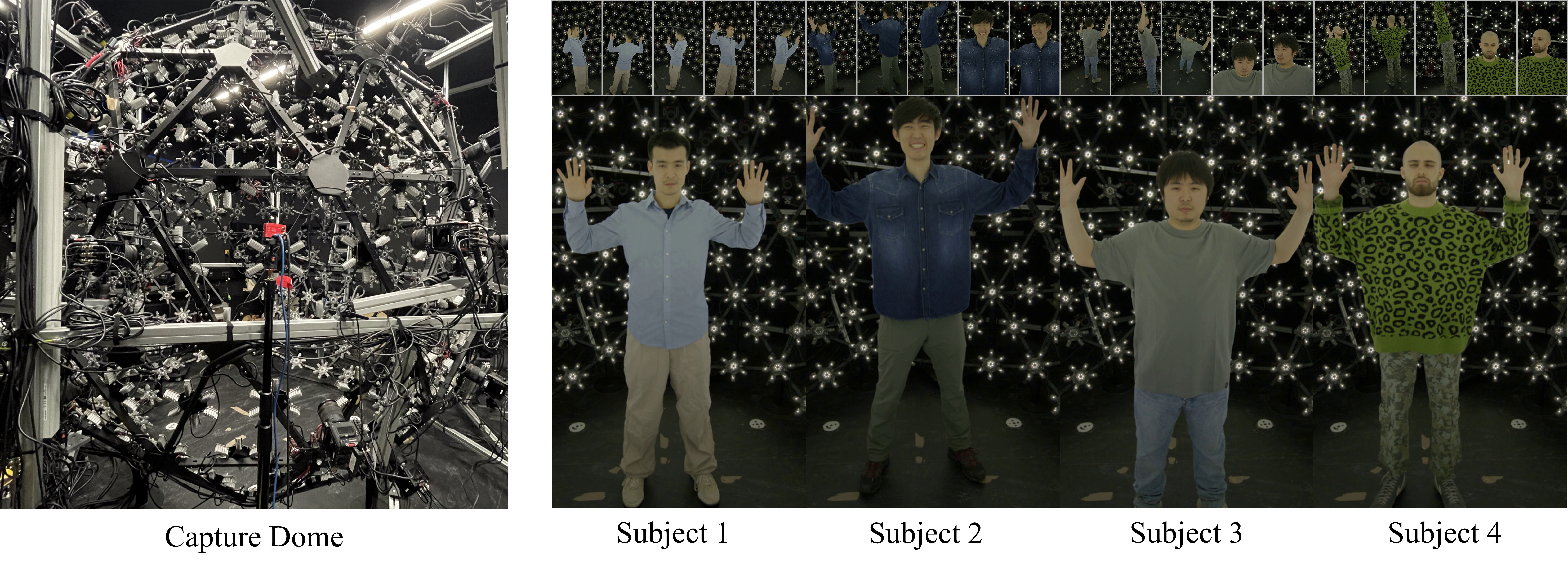}
    \caption{Illustration of our data capture setup and the four collected subjects.}
    \label{fig:supp_datacap}
\end{figure}

%% file: Supp_Sections/02_additional_results.tex
\section{Additional Results}
\label{sec:add_results}

In this section, we give detailed quantitative and additional qualitative results to supplement the evaluations presented in the main paper.

\subsection{Detailed Quantitative Results}
\label{sec:add_quant}

\input{Supp_Tables/supp_quant_baseline_detailed}

We provide detailed quantitative results for the baseline comparison (Sec. 4.3) in~\cref{tab:detailed_quant_baseline_compare}, the ablation on the prior (Sec. 4.4) in~\cref{tab:detailed_quant_ablation_prior}, and the ablation on background element masking and additional enhancement (Sec. 4.4) in~\cref{tab:detailed_quant_ablation_maskprior}.

\input{Supp_Tables/supp_quant_ablation_prior_detailed}

\subsection{Additional Qualitative Results}
\label{sec:add_qual}

\input{Supp_Tables/supp_quant_ablation_maskcouple_detailed}
\input{Supp_Figures/additional_baselines}
\input{Supp_Figures/crossinference_consistency}
\input{Supp_Figures/additional_views}

We provide additional qualitative results: \cref{fig:baselines_02} shows additional qualitative results for the baseline comparison for subjects 3 and 4. \cref{fig:ci_consistency} illustrates inference-dependent shifts present within a single $57$-frame sequence and across multiple separate inference runs for the original non-fine-tuned relighting model and after fine-tuning. Finally, \cref{fig:views_02} provides additional visualizations of view consistency between inferences.

%% file: Supp_Tables/supp_quant_baseline_detailed.tex
\begin{table}[t]
    \centering
    \small
    \setlength{\tabcolsep}{3pt} 
    \caption{Detailed per-subject comparison results against the constructed baselines (Sec 4.3)}
    \begin{tabular}{lcccccc}
        \toprule
        %
        %
        \multirow{2}{*}{\textbf{Subject 1}} 
          & \multicolumn{3}{c}{\textbf{Novel View/Motion}} 
          & \multicolumn{3}{c}{\textbf{Novel View/Motion/Light}} \\
        
        \cmidrule(lr){2-4} 
        \cmidrule(lr){5-7}
        
        & PSNR $\uparrow$ & LPIPS $\downarrow$ & SSIM $\uparrow$
        & PSNR $\uparrow$ & LPIPS $\downarrow$ & SSIM $\uparrow$ \\
        
        \midrule
        MA/PBR~\cite{chen2024meshavatarlearninghighqualitytriangular}
        & 20.51 &0.0860 & 0.910
        & 20.26 &0.0860 &0.908 \\

        MA/DR~\cite{chen2024meshavatarlearninghighqualitytriangular, liang2025DiffusionRenderer} 
        & \underline{23.87} &  \underline{0.0710} &  \underline{0.940} 
        &  \underline{23.46}  &  \underline{0.0710} &  \underline{0.938}  \\

        EVA/IC~\cite{junkawitsch2025eva, zhang2025scaling}
        & 7.41 & 0.0987 & 0.867
        & 7.62 & 0.0969 & 0.871  \\
        
        \textbf{Ours}    
        & \textbf{25.73} & \textbf{0.0659} &  \textbf{0.948}
        & \textbf{25.50} & \textbf{0.0652}  & \textbf{0.948}  \\
        
        \bottomrule\addlinespace[4pt]
        %
        %
        \multirow{2}{*}{\textbf{Subject 2}} 
          & \multicolumn{3}{c}{\textbf{Novel View/Motion}} 
          & \multicolumn{3}{c}{\textbf{Novel View/Motion/Light}} \\
        
        \cmidrule(lr){2-4} 
        \cmidrule(lr){5-7}
        
        & PSNR $\uparrow$ & LPIPS $\downarrow$ & SSIM $\uparrow$
        & PSNR $\uparrow$ & LPIPS $\downarrow$ & SSIM $\uparrow$ \\
        
        \midrule
        MA/PBR~\cite{chen2024meshavatarlearninghighqualitytriangular}
        & 24.53 & 0.123 & 0.877
        & 24.27 & 0.122 & 0.877\\

        \textbf{MA/DR}~\cite{chen2024meshavatarlearninghighqualitytriangular, liang2025DiffusionRenderer} 
        &  \underline{28.69}  & \textbf{0.093} & \underline{0.937}
        & \textbf{28.35} & \textbf{0.092} & \underline{0.936} \\

        EVA/IC~\cite{junkawitsch2025eva, zhang2025scaling}
        & 14.22 & 0.129  & 0.835
        & 14.37  & 0.126 &  0.839 \\
        
        \textbf{Ours}    
        & \textbf{28.72} & \textbf{0.093} &  \textbf{0.939}
        & \underline{28.34}  & \underline{0.094} & \textbf{0.938}  \\
        
        \bottomrule\addlinespace[4pt]
        
        %
        %
        
        \multirow{2}{*}{\textbf{Subject 3}} 
          & \multicolumn{3}{c}{\textbf{Novel View/Motion}} 
          & \multicolumn{3}{c}{\textbf{Novel View/Motion/Light}} \\
        
        \cmidrule(lr){2-4} 
        \cmidrule(lr){5-7}
        
        & PSNR $\uparrow$ & LPIPS $\downarrow$ & SSIM $\uparrow$
        & PSNR $\uparrow$ & LPIPS $\downarrow$ & SSIM $\uparrow$ \\
        
        \midrule
        MA/PBR~\cite{chen2024meshavatarlearninghighqualitytriangular}
        & 22.73 & 0.081 & 0.921
        & 22.54 & 0.080 & 0.921\\

        \textbf{MA/DR}~\cite{chen2024meshavatarlearninghighqualitytriangular, liang2025DiffusionRenderer} 
        & \textbf{27.25} & \textbf{0.064}  & \textbf{0.956} 
        & \textbf{26.98} & \textbf{0.064} & \textbf{0.956} \\

        EVA/IC~\cite{junkawitsch2025eva, zhang2025scaling}
        & 8.89 & 0.102 & 0.869
        & 9.02 & 0.100 &  0.871 \\
        
        \textbf{Ours}    
        & \underline{27.21} & \underline{0.065} &  \underline{0.954}
        & \underline{26.89} & \underline{0.065}  & \underline{0.953}  \\
        
        \bottomrule\addlinespace[4pt]
        
        %
        %
        
        \multirow{2}{*}{\textbf{Subject 4}} 
        & \multicolumn{3}{c}{\textbf{Novel View/Motion}} 
        & \multicolumn{3}{c}{\textbf{Novel View/Motion/Light}} \\
        
        \cmidrule(lr){2-4} 
        \cmidrule(lr){5-7}
        
        & PSNR $\uparrow$ & LPIPS $\downarrow$ & SSIM $\uparrow$
        & PSNR $\uparrow$ & LPIPS $\downarrow$ & SSIM $\uparrow$ \\
        
        \midrule
        MA/PBR~\cite{chen2024meshavatarlearninghighqualitytriangular}
        & 22.68 & \underline{0.090} & 0.889
        & 22.45 & \underline{0.090} & 0.888 \\

        MA/DR~\cite{chen2024meshavatarlearninghighqualitytriangular, liang2025DiffusionRenderer} 
        & \underline{24.49} & 0.092 &  \underline{0.905}
        & \underline{24.14} & 0.093 & \underline{0.903} \\

        EVA/IC~\cite{junkawitsch2025eva, zhang2025scaling}
        & 10.50 & 0.104 & 0.840
        & 10.62 & 0.102  & 0.841  \\
        
        \textbf{Ours}    
        & \textbf{24.82} & \textbf{0.084} &  \textbf{0.906}
        & \textbf{24.44} & \textbf{0.084} & \textbf{0.904}  \\
        
        \bottomrule
    \end{tabular}
    \label{tab:detailed_quant_baseline_compare}

\end{table}

%% file: Supp_Tables/supp_quant_ablation_prior_detailed.tex
\begin{table}[t]
    \centering
    \small
    \setlength{\tabcolsep}{3pt} 
    \caption{Detailed per-subject comparison results for the ablation on the prior (Sec. 4.4 A). GPU hours shown in the main paper are the same across subjects.}
    \begin{tabular}{lcccccc}
        \toprule
        %
        %
         
          & \multicolumn{3}{c}{\textbf{Subject 2}} 
          & \multicolumn{3}{c}{\textbf{Subject 3}} \\
        
        \cmidrule(lr){2-4} 
        \cmidrule(lr){5-7}
        
        & PSNR $\uparrow$ & LPIPS $\downarrow$ & SSIM $\uparrow$
        & PSNR $\uparrow$ & LPIPS $\downarrow$ & SSIM $\uparrow$ \\
        
        \midrule
        \textbf{LoRA DR (Ours)}
        & \textbf{28.34} & \textbf{0.094} & \textbf{0.938}
        & \underline{26.89} & \underline{0.065} & \textbf{0.953} \\

        (a) w/o finetune
        & 20.00 & 0.131 & 0.876
        & 18.35 & 0.109 & 0.903  \\

        (b) with full finetune
        & \underline{28.25} & \underline{0.096} & \textbf{0.938}
        & \textbf{27.51} & \textbf{0.062} & \underline{0.955} \\
        
        (c) w/o prior 
        & 22.10 & 0.114 & 0.896
        & 20.43 & 0.079 & 0.924 \\
        \bottomrule
    \end{tabular}
    
    \label{tab:detailed_quant_ablation_prior}    
\end{table}

%% file: Supp_Tables/supp_quant_ablation_maskcouple_detailed.tex
\begin{table}[t]
    \centering
    \small
    \setlength{\tabcolsep}{3pt} 
    \caption{Detailed per-subject comparison results for the ablation on additional enhancement and background masking (Sec 4.4 B+C). Please view the main paper for a detailed discussion.}
    \begin{tabular}{lcccccc}
        \toprule
        %
        %
         
          & \multicolumn{3}{c}{\textbf{Subject 1}} 
          & \multicolumn{3}{c}{\textbf{Subject 2}} \\
        
        \cmidrule(lr){2-4} 
        \cmidrule(lr){5-7}
        
        & PSNR $\uparrow$ & LPIPS $\downarrow$ & SSIM $\uparrow$
        & PSNR $\uparrow$ & LPIPS $\downarrow$ & SSIM $\uparrow$ \\
        
        \midrule
        \textbf{Relight only (Ours)}
        & \textbf{25.50} & 0.065 & \textbf{0.948}
        & 28.34 & 0.094 & 0.938 \\

        Relight + Enhance
        & 24.39 & \textbf{0.062} & 0.947
        & \textbf{28.84} & \textbf{0.087} & \textbf{0.943} \\

        \textbf{w/o BG masks  (Ours)}
        & 25.50 & 0.065 & 0.948
        & 28.35 & 0.094 & 0.938\\
        
        w/ BG masks
        & \textbf{25.94} & \textbf{0.062} & \textbf{0.950}
        & \textbf{28.74} & \textbf{0.088} & \textbf{0.943}\\
        \bottomrule\addlinespace[4pt]

          & \multicolumn{3}{c}{\textbf{Subject 3}} 
          & \multicolumn{3}{c}{\textbf{Subject 4}} \\
        
        \cmidrule(lr){2-4} 
        \cmidrule(lr){5-7}
        
        & PSNR $\uparrow$ & LPIPS $\downarrow$ & SSIM $\uparrow$
        & PSNR $\uparrow$ & LPIPS $\downarrow$ & SSIM $\uparrow$ \\
        
        \midrule
        \textbf{Relight only (Ours)}
        & \textbf{26.89} & 0.065  & 0.953
        & 24.44 & 0.084 & 0.904 \\

        Relight + Enhance
        & 26.80 & \textbf{0.062} & \textbf{0.955}
        & \textbf{24.46} & \textbf{0.077} & \textbf{0.905}\\
        
        \textbf{w/o BG masks  (Ours)}
        & 26.89 & 0.065  & 0.953
        & 24.44 & 0.084 & 0.904 \\
        
        w/ BG masks
        & \textbf{27.99} & \textbf{0.061} & \textbf{0.958}
        & \textbf{24.63} & \textbf{0.083} & \textbf{0.906}\\
        \bottomrule\addlinespace[4pt]
    \end{tabular}    
    \label{tab:detailed_quant_ablation_maskprior}    
\end{table}

%% file: Supp_Figures/additional_baselines.tex
\begin{figure}
    \centering
    \includegraphics[width=0.95\textwidth]{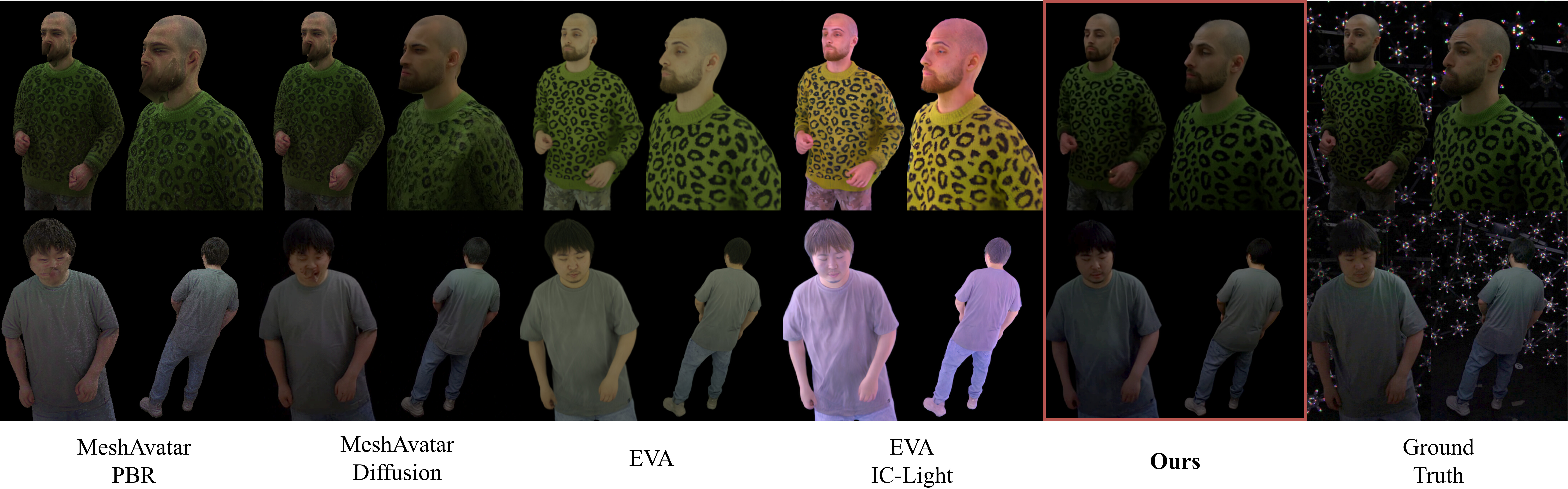}
    \caption{Additional qualitative results for the comparison against our constructed baselines (Sec 4.3).}
    \label{fig:baselines_02}
\end{figure}

%% file: Supp_Figures/crossinference_consistency.tex
\begin{figure}
    \centering
    \includegraphics[width=0.95\textwidth]{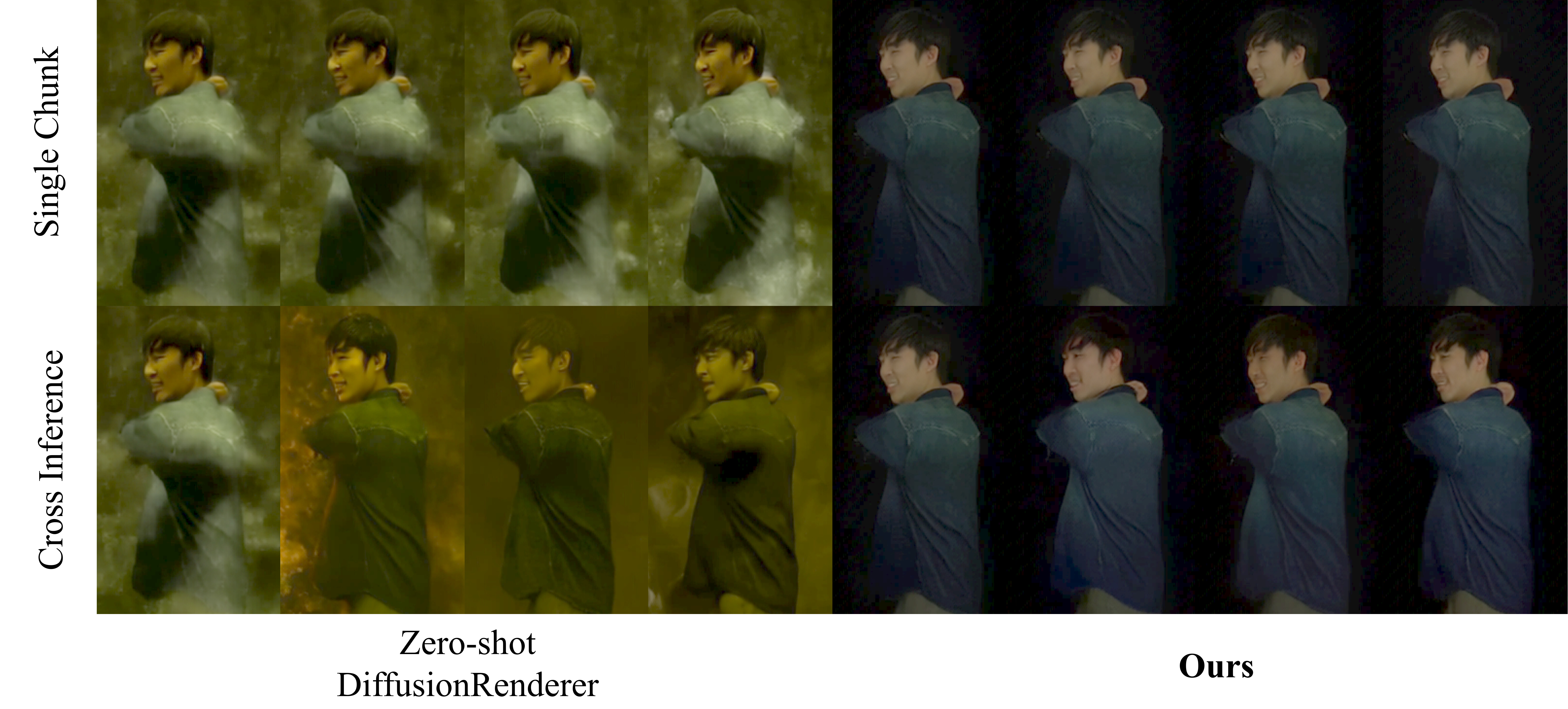}
    \caption{Illustration of illumination consistency for a single rendered chunk (above) and across separate inference runs (below) for the non-finetuned (left) and our LoRA fine-tuned model (right). As can be seen, both models are temporally consistent within a single chunk. However, the non-finetuned version exhibits significantly larger variations across inferences than the fine-tuned version.}
    \label{fig:ci_consistency}
\end{figure}

%% file: Supp_Figures/additional_views.tex
\begin{figure}
    \centering
    \includegraphics[width=0.95\textwidth]{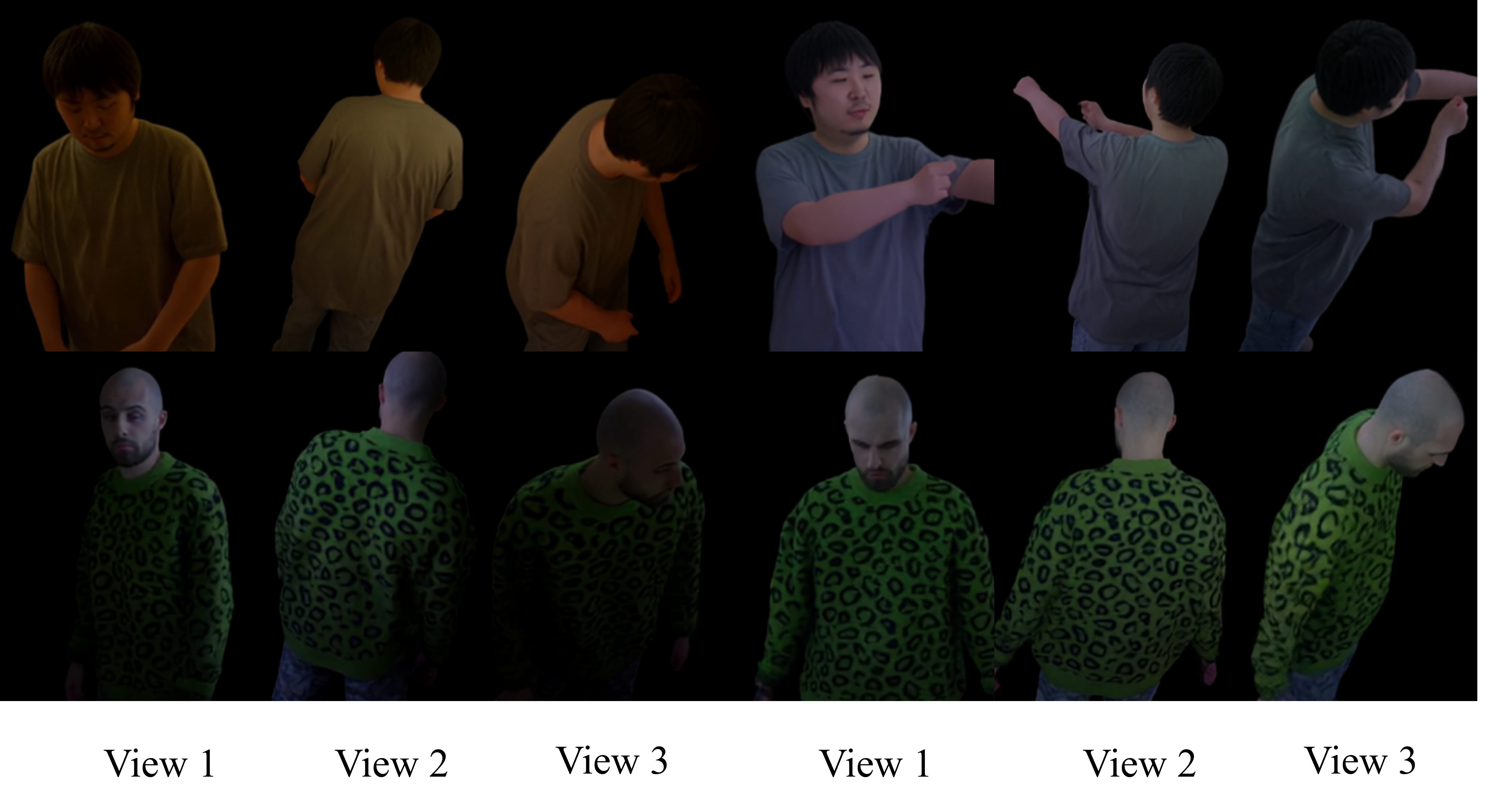}
    \caption{Additional visualization of view consistency across separate inference runs, as shown in Fig. 7 of the main paper. Please note the that left and right views are showing a separate light condition.}
    \label{fig:views_02}
    \vspace{-15pt}
\end{figure}

%% file: Supp_Sections/03_generalization.tex
\section{Cross-subject Generalization}
\label{sec:generalization}

We provide an additional qualitative cross-subject generalization experiment across subjects $2$, $3$, and $4$. To this end, we LoRA fine-tune a diffusion relighting model for every true subset of subjects, then test their performance across all subjects. We note that subject $1$ is excluded from this evaluation, as they were recorded in a separate capture session and do not have lighting consistent with the other subjects (\cref{sec:supp_data}).
As can be seen in~\cref{fig:ablation_generalization}, although color shifts between the subject-specific and non-subject-specific models remain noticeable, the non-subject-specific models produce plausible results, even modeling complex lighting effects such as specularities in the hair. As such, generalization of the relighting model appears as a promising future research direction. Future work will include collecting a larger set of diverse subjects to evaluate generalization capability conclusively.

%% file: Supp_Sections/05_shift.tex
\section{Influence of Misaligned Frame Pairs}
\label{sec:misalignment}

We note that as our subjects are actively moving during our captures, the flat-lit and relit frame pairs are not perfectly aligned. Consequently, as shown in~\cref{fig:shift} sufficiently strong motion (Subject 2, average $\sim12.18 \text{mm}$ per frame) causes the LoRA fine-tuned relighting model to slightly shift the subject around to correct for the misalignment. We observe that training on a slower subject (Subject 3, average $\sim8.84 \text{mm}$ per frame) correspondingly results in overall less shifts for the final model. Interestingly, despite this contamination, the main structure of the subject --- including details in the face and main body --- remains typically unaffected. However, as also shown in~\cref{fig:shift}, failure cases induced by the misalignment are possible. This includes preservation of detail in hands, sometimes producing a blurred out prediction and, very rarely, complete failure for highly dynamic clothing. Future work will explore various methods for additional misalignment compensation, including adjusted data collection schemes and motion-dependent loss weights during training.


\input{Supp_Figures/shift_vis}

%% file: Supp_Figures/shift_vis.tex
\begin{figure}
    \centering
    \includegraphics[width=0.95\textwidth]{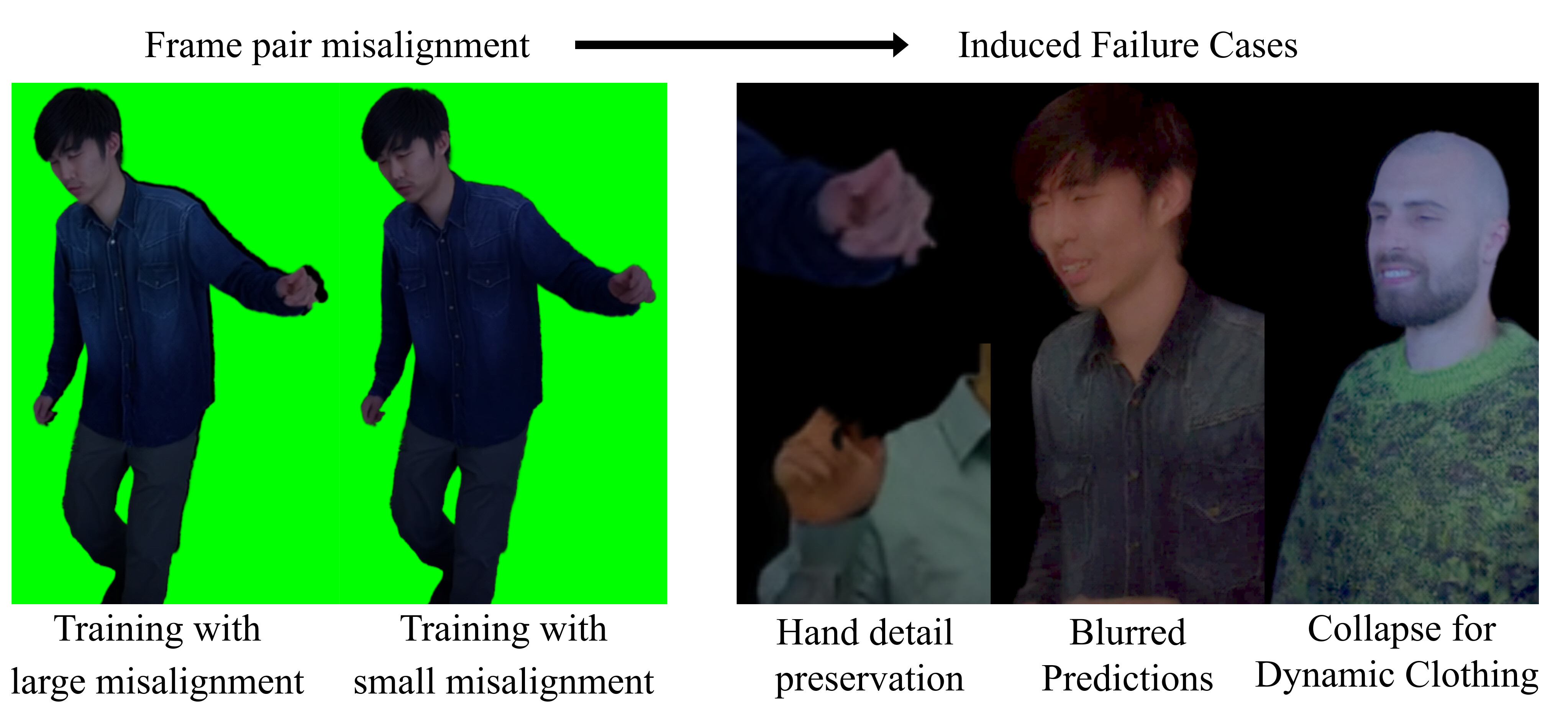}
    \caption{Qualitative demonstration of artifacts caused by misalignment between frame pairs. On the left, it can be seen that larger motion between frame pairs causes the model to slightly shift the subject around, resulting in misalignment with the original EVA mask (green). This effects disappears when training using only a low-motion subject.
    On the right, various rare artifacts induced by the misalignment are shown.}
    \label{fig:shift}
\end{figure}

%% file: Supp_Sections/06_ethical_discussion.tex
\section{Ethical Concerns}
\label{sec:ethics}
The data used to fine-tune the relighting model and train the albedo avatar in our method has been sourced ethically. Every subject was informed about and consented to the use of their captures in the development of our method. Moreover, all subjects agreed to make the captures publicly available for scientific purposes. 

We acknowledge that while the proposed framework opens various opportunities --- i.e. in movie and game production, immersive VR/AR applications for training medical students, and telepresence for remote business/ personal meetings ---  it could also be utilized in the production of nefarious misleading media. While the method currently requires actor-specific data --- hindering quick and easy adoption --- future developments in generative relighting targeting generalization could heighten the potential for misuse, necessitating additional guard rails to ensure the framework is employed ethically.

\input{Main_Figures/ablation_generalization}

%% file: Main_Figures/ablation_generalization.tex
\begin{figure}
    \centering
    \includegraphics[width=\textwidth]{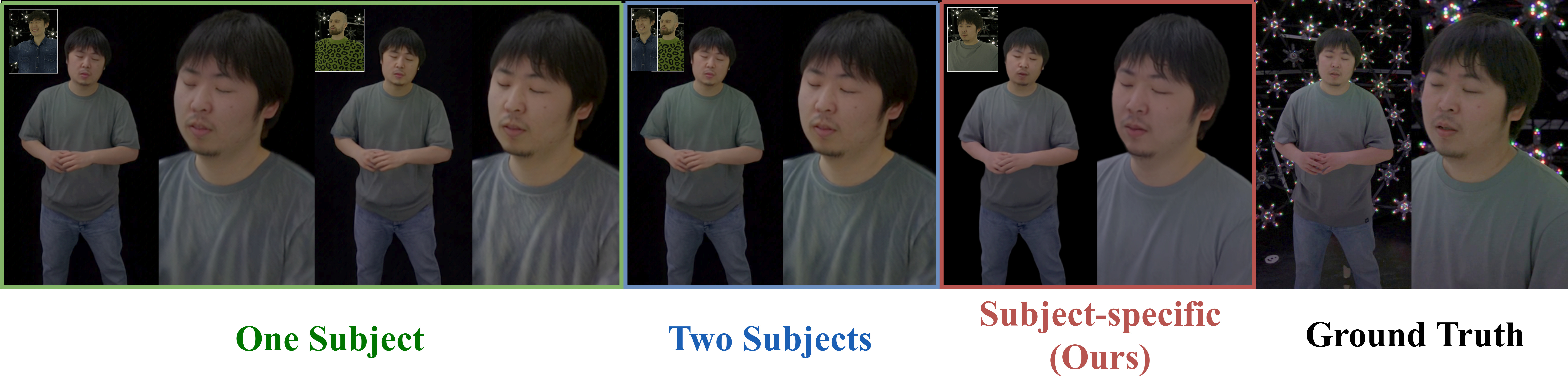}
    \caption{Qualitative illustration of the cross-subject training experiment. The subjects used for training are depicted in the upper left corners. While the non subject-specific models are able to reasonably relight, color shifts compared to subject-specific model remain visible.}
    \label{fig:ablation_generalization}
    \vspace{-10pt}
\end{figure}